\documentclass[prl,twocolumn,showpacs,amsmath,amssymb,superscriptaddress]{revtex4-1}
\usepackage{epsfig,epstopdf} 
\usepackage{graphicx}
\usepackage{dcolumn}
\usepackage{bm}
\usepackage{bbm}
\usepackage{ulem}
\usepackage{color}
\usepackage{slashed}
\usepackage{hyperref}
\usepackage{mathrsfs}
\usepackage{subfigure}
\usepackage{verbatim}
\usepackage{dsfont}
\usepackage{float}

\newcommand{\nc}{\newcommand}
\nc{\be}{\begin{equation}} \nc{\ee}{\end{equation}}
\nc{\bea}{\begin{eqnarray}} \nc{\eea}{\end{eqnarray}}
\nc{\bean}{\begin{eqnarray*}} \nc{\eean}{\end{eqnarray*}}
\nc{\dg}{\dagger}
\nc{\ua}{\uparrow} \nc{\da}{\downarrow}
\nc{\lag}{\langle} \nc{\rag}{\rangle}

\begin{document}

\bibliographystyle{apsrev4-1}

\title{M\"{o}bius Insulator and Higher-Order Topology in MnBi$_{2n}$Te$_{3n+1}$}
\author{Rui-Xing Zhang}
\email{ruixing@umd.edu}
\affiliation{Condensed Matter Theory Center and Joint Quantum Institute, Department of Physics, University of Maryland, College Park, Maryland 20742-4111, USA}
\author{Fengcheng Wu}
\email{wufcheng@umd.edu}
\affiliation{Condensed Matter Theory Center and Joint Quantum Institute, Department of Physics, University of Maryland, College Park, Maryland 20742-4111, USA}
\author{S. Das Sarma}
\affiliation{Condensed Matter Theory Center and Joint Quantum Institute, Department of Physics, University of Maryland, College Park, Maryland 20742-4111, USA}

\begin{abstract}
	We propose MnBi$_{2n}$Te$_{3n+1}$ as a magnetically tunable platform for realizing various symmetry-protected higher-order topology. Its canted antiferromagnetic phase can host exotic topological surface states with a M\"obius twist that are protected by nonsymmorphic symmetry. Moreover, opposite surfaces hosting M\"obius fermions are connected by one-dimensional chiral hinge modes, which offers the {\it first} material candidate of a higher-order topological M\"obius insulator. We uncover a general mechanism to feasibly induce this exotic physics by applying a small in-plane magnetic field to the antiferromagnetic topological insulating phase of MnBi$_{2n}$Te$_{3n+1}$, as well as other proposed axion insulators. For other magnetic configurations, two classes of inversion-protected higher-order topological phases are ubiquitous in this system, which both manifest gapped surfaces and gapless chiral hinge modes. We systematically discuss their classification, microscopic mechanisms, and experimental signatures. Remarkably, the magnetic-field-induced transition between distinct chiral hinge mode configurations provides an effective ``topological magnetic switch". 
\end{abstract}

\date{\today}
\maketitle

\begin{figure}[t]
	\centering
	\includegraphics[width=0.49\textwidth]{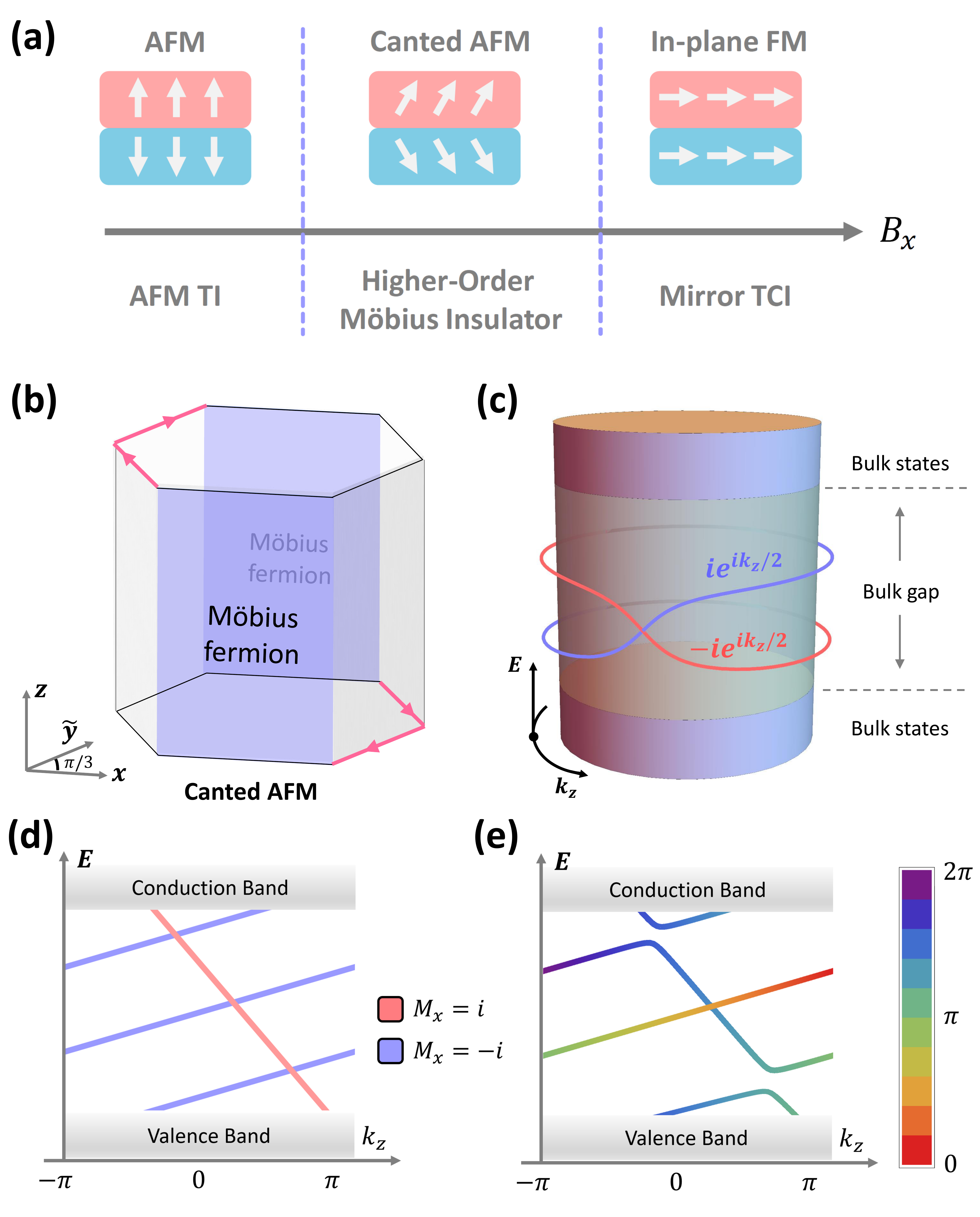}
	\caption{(a) Applying an in-plane field $B_x$ to an AFM TI induces a topological transition to a higher-order M\"obius insulator, and eventually to a mirror TCI protected by $M_x$. (b) Schematic of a higher-order M\"obius insulator in a hexagonal geometry. The red arrows represent chiral hinge modes. (c) Schematic of surface M\"obius fermions along the glide-invariant line (GIL) labeled with a ${\cal G}_x$ eigenvalue of $\pm i e^{ik_z/2}$. Along GIL, the M\"obius fermions can be disconnected from other bands. We schematically show how the surface state of a mirror TCI in (d) evolves into a M\"obius fermion in (e), when the mirror symmetry is broken to ${\cal G}_x$. The color bar in (e) shows the phase of the corresponding ${\cal G}_x$ eigenvalue. A gap opens when two surface states with the same color cross.}
	\label{Fig: Schematic}
\end{figure}

{\it Introduction} - The past decade has witnessed the rapid development of topological crystalline insulators (TCI) as a new class of materials \cite{fu2011topological,hsieh2012topological,mong2010antiferro,ando2015topological,liu2014topological,zhang2015topological,shiozaki2015z2,fang2015new,wang2016hourglass,chang2017mobius}, where crystalline symmetries protect band topology in solids. The bulk topology of a TCI enforces protected in-gap states to emerge only on its symmetry-preserving boundaries. Recently, it was realized that a special class of TCI also features higher-order topology \cite{benalcazar2017quantized,benalcazar2017electric,schindler2018higher,langbehn2017reflection,khalaf2018higher,khalaf2018symmetry,zhang2013surface}, where gapless modes live on $(D-d)$-dimensional boundary of a $D$-dimensional TCI with $1<d\leq D$. Theoretical work on these higher-order topological insulators (HOTI) has mainly focused on topological classifications and model constructions, with only a few realistic candidate materials being proposed \cite{schindler2018Bi,xu2019higher,yue2019symmetry,wang2018higher,lee2019higher,sheng2019two}. Experimentally, the only evidence for electronic HOTI was demonstrated in bismuth \cite{schindler2018Bi}. Therefore, identifying more experimentally accessible HOTI systems is important.

Recently, a major breakthrough for TCI is the discovery of antiferromagnetic (AFM) topological insulators (TI) in MnBi$_{2n}$Te$_{3n+1}$ family of materials \cite{zhang2019MBT,otrokov2018prediction,gong2019experimental,li2019intrinsic,vidal2019surface,hu2019van,wu2019natural,li2019dirac,hao2019gapless,chen2019topological}. With intrinsic $A$-type AFM order and an out-of-plane easy axis, the band topology of MnBi$_{2n}$Te$_{3n+1}$ is protected by an AFM time-reversal symmetry (TRS) $\Theta_M$, which combines the TRS operation $\Theta$ and a half-unit-cell translation $T_{[00\frac{1}{2}]}$ along $\hat{z}$ direction. Compounds with $n=1,2,3$ (i.e., MnBi$_2$Te$_4$, MnBi$_4$Te$_7$, and MnBi$_6$Te$_{10}$) are currently under active experimental study \cite{hu2019van,wu2019natural,vidal2019topological,shi2019magnetic,ding2019crystal,tian2019magnetic,xu2019persistent,hu2019universal,lee2019spin,yan2019atype,yan2019crystal}. Remarkably, evidence of quantum anomalous Hall effect \cite{deng2019magnetic,ge2019highchernnumber} and axion insulator \cite{liu2019quantum} has recently been reported in few-layer MnBi$_2$Te$_4$.  Since the magnetic moments of MnBi$_{2n}$Te$_{3n+1}$ can be easily manipulated by a weak applied magnetic field, it is an interesting open question on the type of band topology that could arise for various field-induced magnetic configurations in MnBi$_{2n}$Te$_{3n+1}$.

In this Letter, we propose MnBi$_{2n}$Te$_{3n+1}$ as a highly tunable system to realize a variety of HOTI phases. 
In particular, we show that applying an in-plane magnetic field cants the AFM ordering and leads to the {\it first} material platform for a higher-order M\"obius insulator with a M\"obius twist in its topological surface state, as schematically shown in Fig. \ref{Fig: Schematic}. Furthermore, opposite surfaces hosting M\"obius states are connected by 1d chiral hinge modes, manifesting the higher-order nature. For general magnetic configurations, two distinct classes of inversion-protected HOTIs are expected in MnBi$_{2n}$Te$_{3n+1}$. These two HOTI phases share the same bulk topological index but differ in their hinge mode configurations. 
Rotating the magnetic field can drive transition between these two phases, and therefore, can lead to a topological magnetic switching of $G_{zz}$, the two-terminal conductance along $\hat{z}$ direction. We also discuss experimental consequences and application of our theory to other proposed axion insulators.   

{\it Model Hamiltonian} - 
We start by defining an effective Hamiltonian for MnBi$_{2n}$Te$_{3n+1}$ that captures its essential symmetry and topological features. In the absence of magnetism, the point group of MnBi$_{2n}$Te$_{3n+1}$ is $D_{3d}$ \cite{footnote1}, which can be generated by (i) a three-fold rotation $C_{3z}$ around $z$-axis, (ii) a two-fold rotation $C_{2x}$ around $x$-axis, and (iii) the spatial inversion ${\cal I}$. $D_{3d}$ also contains three in-plane mirror operations including $M_x$. Following earlier first-principles calculations \cite{zhang2019MBT,vidal2019topological}, we consider the basis functions $|\uparrow(\downarrow), \pm\rangle$ with $\pm$ parity eigenvalues. This defines a four-band ${\bf k}\cdot {\bf p}$ Hamiltonian $H_0({\bf k})$ around $\Gamma$ point, which resembles that for Bi$_2$Se$_3$ \cite{liu2010model} and describes a 3d massive Dirac fermion. In particular, $H_0 ({\bf k}) = e({\bf k}) \mathbb{I}_4 + \sum_{i=1}^5 d_i({\bf k}) \Gamma_i$,
where $\mathbb{I}_4$ is the identity matrix, $\Gamma_i = s_i\otimes \sigma_1$ for $i=1,2,3$, $\Gamma_4=s_0\otimes \sigma_2$, and $\Gamma_5 = s_0\otimes \sigma_3$. $s_i$ and $\sigma_i$ are the Pauli matrices for the spin and orbital degrees of freedom, respectively. The point group symmetry constrains the explicit forms of $d_i({\bf k})$ to be
$d_1 = vk_x,\ d_2 = vk_y,\ d_3 = v_z k_z,\ d_4 = w (k_+^3 + k_-^3),\ d_5 = M_0 + M_1 k_z^2 + M_2 (k_x^2+k_y^2)$, and $e({\bf k}) = C_0 + C_1 k_z^2 + C_2 (k_x^2+k_y^2)$. Physically, $v$ and $v_z$ denote the in-plane and out-of-plane Fermi velocities, while $w$ controls the hexagonal warping effect and reduces the full rotation symmetry down to $C_{3z}$. 
For our purpose, we regularize our model on a 3d hexagonal lattice as shown in Fig. \ref{Fig: AFM} (a). The full expression for the lattice model is given in the Supplemental Material (SM) \cite{supplementary}.

The pristine MnBi$_{2n}$Te$_{3n+1}$ compounds usually develop A-type AFM ordering along (001) direction, as shown in Fig. \ref{Fig: AFM} (a), where we introduce a sublayer index $i=A,B$ to describe the AFM-induced unit cell doubling. We characterize the magnetization in sublayer $i$ by ${\bf M}_i=(\cos \phi_i \sin\theta_i, \sin\phi_i \sin\theta_i, \cos\theta_i)$ with angles $\phi_i$ and $\theta_i$. In particular, the exchange coupling term is
\begin{equation}
	H_\text{ex} = \begin{pmatrix}
	{\bf M}_A \cdot \vec{s}\otimes \sigma_0 & 0 \\
	0 & {\bf M}_B \cdot \vec{s}\otimes \sigma_0 \\
	\end{pmatrix}.
\end{equation}
The AFM ordering is described by $(\theta_A,\theta_B)=(0,\pi)$. With a set of band parameters that captures the bulk band inversion at $\Gamma$ \cite{supplementary}, we calculate the energy spectrum for the (010) surface of the AFM phase in a semi-infinite geometry using iterative Green function method. As shown in Fig. \ref{Fig: AFM} (b), the system hosts a single gapless Dirac cone at $\bar{\Gamma}$, the origin of surface Brillouin zone (BZ). The gapless Dirac surface state here is protected by the AFM TRS $\Theta_M$ and generally shows up for any surface that is compatible with $\Theta_M$ symmetry \cite{mong2010antiferro}. On the other hand, a finite surface energy gap is expected on the (001) surface. The AFM TI phase serves as the starting point for our discussion on higher-order topology.

\begin{figure}[t]
	\centering
	\includegraphics[width=0.45\textwidth]{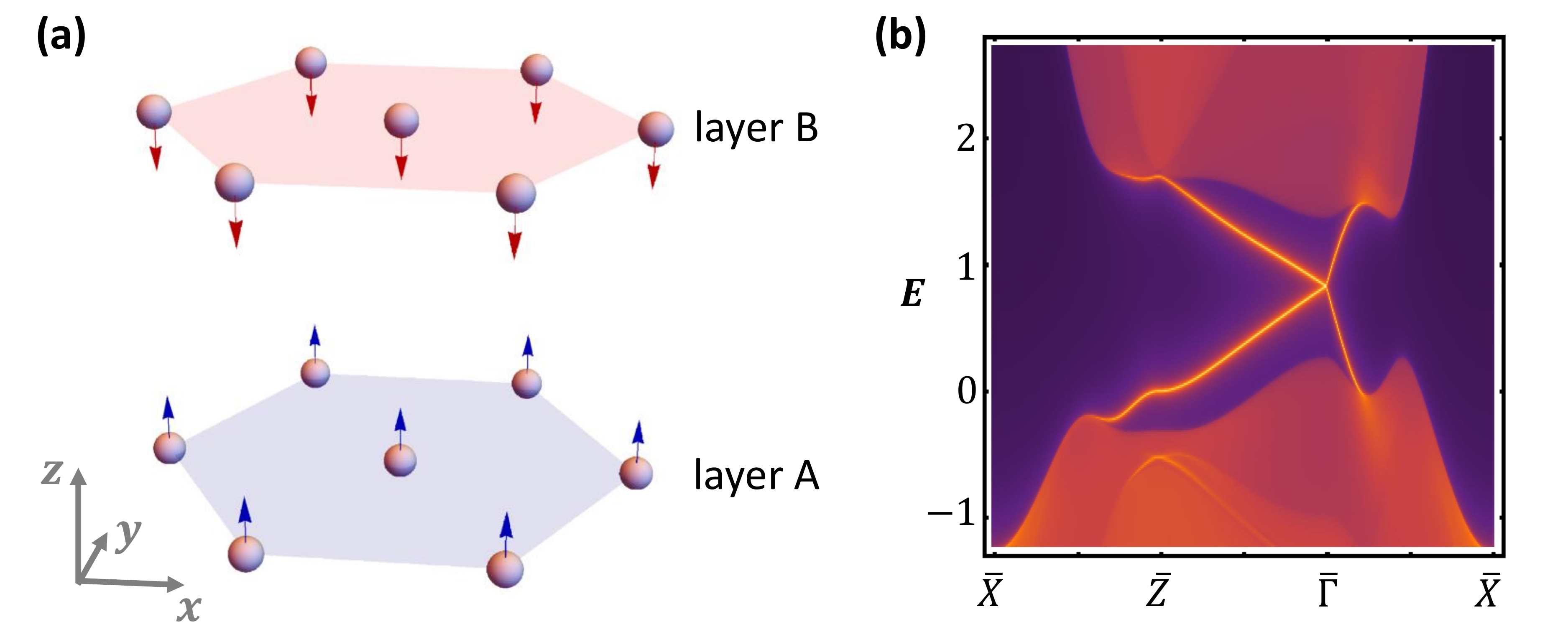}
	\caption{ (a) Lattice structure of our tight-binding model. (b) Surface spectrum for (010) surface of the AFM phase.}
	\label{Fig: AFM}
\end{figure}

{\it Higher-Order M\"obius Insulator} - In the presence of an external in-plane magnetic field ${\bf B}$, the AFM cants along the field direction [see Fig. \ref{Fig: Schematic} (a) and discussions in the SM \cite{supplementary}] and generally breaks all symmetries except for the spatial inversion ${\cal I}$. The loss of $\Theta_M$ generally leads to a magnetic surface gap for every surface and trivializes the $\Theta_M$-protected topology. However, this enables the possibility of higher-order topology emerging in this system. 

\begin{figure}[t]
	\centering
	\includegraphics[width=0.48\textwidth]{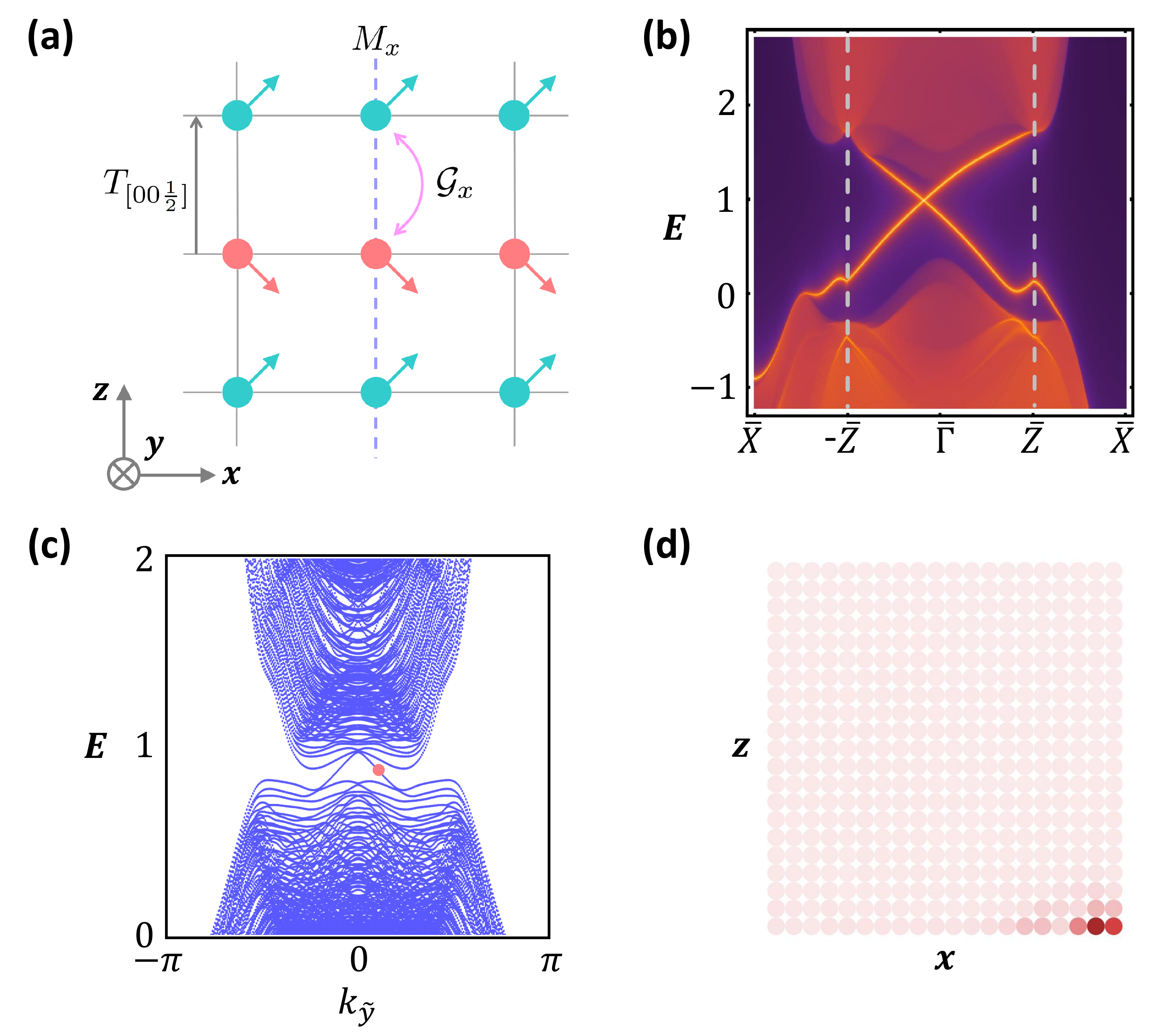}
	\caption{(a) When magnetic moments in the canted AFM cant towards $\hat{x}$ direction, the system has a glide mirror symmetry ${\cal G}_x$, which interchanges the green and red magnetic moments. (b) The surface spectrum on (010) surface for the canted AFM phase with $\theta_A=\pi-\theta_B = 0.3\pi$ and $\phi_A=\phi_B=0$. The glide-invariant line is sanwiched between two white dashed lines, where the M\"obius fermion is well-defined. (c) The corresponding in-gap chiral hinge modes in a prism geometry periodic in $\tilde{y}$ direction. (d) The real-space distribution in the $x$-$z$ plane for the chiral modes marked by the red dot in (c). }
	\label{Fig: Mobius}
\end{figure}

When ${\bf B}$ is along $\hat{x}$ and hence perpendicular to the $M_x$ mirror plane, the canted AFM ordering respects a non-symmorphic glide mirror symmetry ${\cal G}_x$ that combines mirror reflection $M_x$ and the half-unit-cell translation $T_{[00\frac{1}{2}]}$ along $\hat{z}$, as shown in Fig. \ref{Fig: Mobius} (a). Along the glide-invariant line (GIL) of the nonsymmorphic 010 surface BZ (e.g. $k_x=0,\pi$), the surface states are labeled by their glide eigenvalues $g_x = \pm i e^{ik_z/2}$. Whenever surface states with distinct $g_x$ cross, a ``locally" robust surface Dirac point is formed. Nevertheless, only an odd number of surface Dirac points is {\it topologically robust} for this system, featuring $\mathbb{Z}_2$ band topology \cite{fang2015new,shiozaki2015z2}. This motivates us to calculate the (010) surface spectrum for our canted AFM system. As shown in Fig. \ref{Fig: Mobius} (b), a single surface Dirac point is clearly revealed along GIL, which proves the topological nature of the canted AFM phase.

While the crystal momentum is $2\pi$ periodic along GIL, the glide eigenvalue $g_x$ follows a periodicity of $4\pi$ due to the half-unit-cell translation. Therefore, the surface state manifold along GIL manifests itself as a M\"obius twist for $g_x$ \cite{shiozaki2015z2,wang2016hourglass,chang2017mobius,wieder2018wallpaper}, as schematically plotted in Fig. \ref{Fig: Schematic} (c). We thus dub this surface state as 2d ``{\it M\"obius fermions}". Distinct from conventional surface states of topological insulators, the M\"obius fermions can be disconnected from other surface or bulk bands along GIL, but are connected to higher-energy bands away from GIL for being unremovable. In our canted AFM phase, M\"obius fermions only live on the (010) surfaces [i.e., the purple surfaces in Fig. \ref{Fig: Schematic} (b)] where glide symmetry is preserved, while surface gaps show up on all other surfaces. The topological characterization of the M\"obius fermions using Wilson loop method is discussed in the SM \cite{supplementary}.

Remarkably, there exist {\it hinge-localized 1d chiral modes} that connect the opposite (010) surfaces with M\"obius fermions, as shown in Fig. \ref{Fig: Schematic} (b). Indeed, while the glide symmetry ${\cal G}_x$ allows for a $\mathbb{Z}_2$ invariant $\nu_g$ to characterize the M\"obius fermions, a $\mathbb{Z}_4$ symmetry indicator $\kappa$ for the inversion symmetry ${\cal I}$ can be simultaneously defined as \cite{turner2012quantized,ono2018unified}
\bea
\kappa \equiv \sum_{k_i} \frac{(n_+ - n_-)}{2}\ \ (\text{mod }4).
\eea     
Here $n_{\pm}$ counts the number of occupied bands with $\pm$ parity eigenvalues and the summation is over all inversion-symmetric crystal momenta. Physically, an odd $\kappa$ implies a Weyl semimetal. When a system is known to be gapped, the $\kappa = 2$ phase is an axion insulator with higher-order topology \cite{wieder2018axion}. Crucially, symmetry argument imposes a relation between $\nu_g$ and $\kappa$ as \cite{kim2019glide}
\bea
\nu_g \equiv \frac{\kappa}{2}\ \ \text{mod }2.
\eea
As a result, when M\"obius fermions show up ($\nu_g=1$), the chiral hinge mode is required to appear because of $\kappa=2$, and vice versa. 

\begin{figure*}[t]
	\centering
	\includegraphics[width=0.8\textwidth]{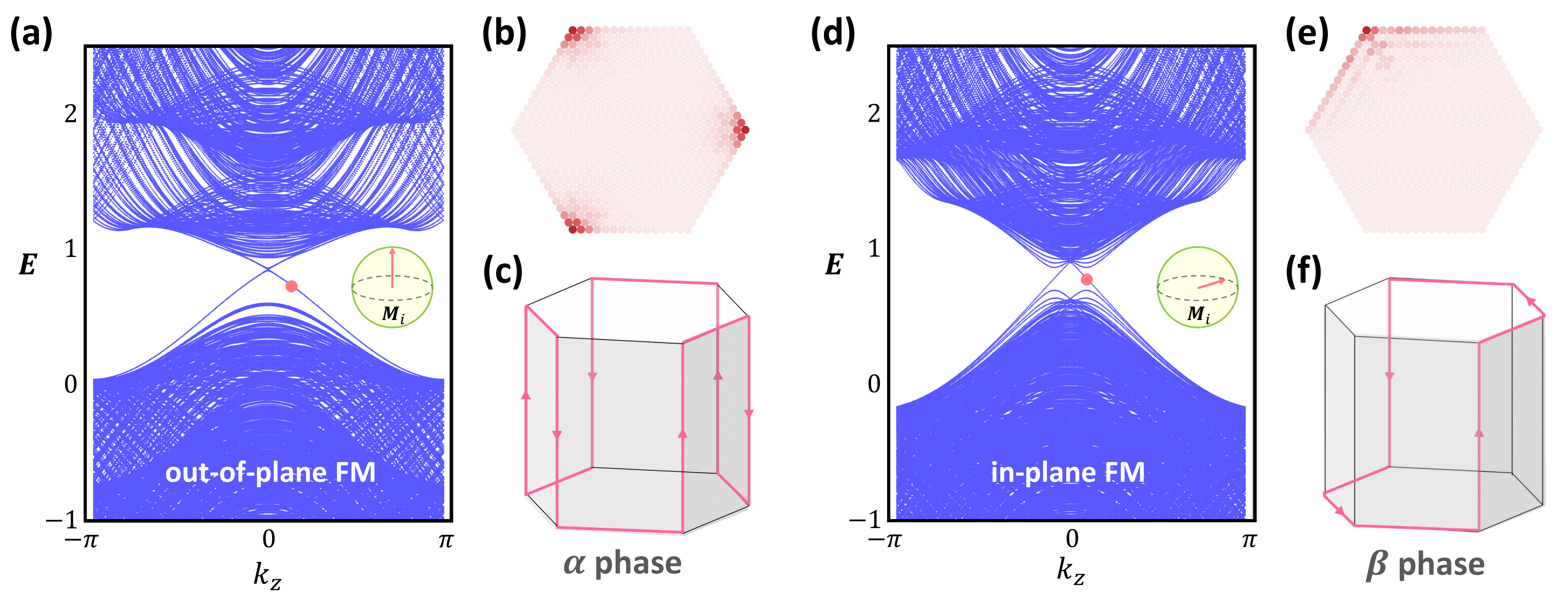}
	\caption{(a) The hinge mode spectrum of the HOTI $\alpha$ phase with $\theta_A=\theta_B=0$. (b) The spatial profile of the left-moving hinge modes indicated by the red dot in (a), which confirms the schematic plot in (c). (d)-(f) Corresponding plots for the HOTI $\beta$ phase. $\phi_A=\phi_B=\frac{\pi}{6}$ and $\theta_A=\theta_B=\frac{\pi}{2}$ for calculations done in (d) and (e).}
	\label{Fig: FM}
\end{figure*}

A simple topological index analysis in the SM \cite{supplementary} implies $\kappa=2$ in our model if the transition from AFM to canted AFM does NOT close the bulk gap. Numerically, we consider a prism geometry periodic along $\tilde{y}$ with a finite cross section in its $x$-$z$ plane ($20\times 21$ lattice sites). Note that $\tilde{y}$ differs from the Cartesian-coordinate $\hat{y}$ axis  by a $\pi/6$ rotation around $\hat{z}$. Fig. \ref{Fig: Mobius} (c) plots the energy spectrum in this prism geometry, showing 1d chiral modes that traverse the surface gaps. In Fig. \ref{Fig: Mobius} (d), we depict the spatial profile of the left-moving mode in the $x$-$z$ cross section and find it localized at the bottom right corner, 
which verifies the hinge-mode picture. 
The existence of  M\"obius fermions on (010) surface and 1d chiral hinge modes along  $\tilde{y}$ direction together establishes our canted AFM phase as a ``{\it higher-order M\"obius insulator}". 

By gradually increasing the in-plane ${\bf B}$ field, the canted AFM phase eventually evolves to an in-plane FM phase, which promotes the glide symmetry ${\cal G}_x$ to a mirror symmetry $M_x$. As shown in Fig. \ref{Fig: Schematic} (a), the higher-order M\"obius phase thus evolves to a mirror-protected TCI \cite{hsieh2012topological,vidal2019topological} with no hinge physics. The transition from a M\"obius fermion to a mirror-protected topological surface state is schematically shown in Figs.~\ref{Fig: Schematic}(d) and \ref{Fig: Schematic}(e).

{\it Inversion-Protected Higher-Order Topology} - When ${\bf B}$ deviates from $\hat{x}$, various new magnetic configurations can be induced that generally break the glide symmetry ${\cal G}_x$ and thus spoil the M\"obius physics, as well as the mirror TCI physics. Despite energy gaps on all surfaces because of lack of symmetries \cite{supplementary}, the inversion indicator $\kappa=2$ remains well-defined as long as the bulk gap survives, which characterizes our system as a robust inversion-symmetric higher-order TI \cite{khalaf2018higher,khalaf2018symmetry} with chiral hinge mode. In particular, there exist two classes of inversion-symmetric HOTI phases in MnBi$_{2n}$Te$_{3n+1}$, which we discuss next. 

When MnBi$_{2n}$Te$_{3n+1}$ is ferromagnetic along $\hat{z}$, it realizes a HOTI that preserves three-fold rotation symmetry $C_{3z}$. When placed on a hexagonal geometry, the hinge mode configuration compatible with both inversion ${\cal I}$ and $C_{3z}$ is shown in Fig. \ref{Fig: FM} (c), which consists of three spatially separated chiral hinge modes along $+\hat{z}$ along with their inversion partners on the opposite hinges. We call this phase the {\it HOTI $\alpha$ phase} to distinguish it from the $\beta$ phase defined later. Numerically, we calculate the energy spectrum of this out-of-plane ferromagnet in the hexagonal prism geometry with in-plane open boundary conditions and with periodicity in $\hat{z}$ direction. As shown in Fig. \ref{Fig: FM} (a), we find three pairs of 1d modes that traverse the surface gap, which are further confirmed as chiral hinge modes by their spatial profiles plotted in Fig. \ref{Fig: FM} (b). These results together confirm the schematic in Fig. \ref{Fig: FM} (c).

In contrast, the {\it HOTI $\beta$ phase} is defined by the chiral hinge mode trajectory shown in Fig. \ref{Fig: FM} (f), which explicitly breaks $C_{3z}$ and has only one pair of chiral hinge modes along $\hat{z}$. The $\beta$ phase can be induced by applying a magnetic field off the high-symmetry directions, which holds for generic magnetic structures in MnBi$_{2n}$Te$_{3n+1}$. An example of $\beta$ phase with FM ordering is numerically confirmed in Fig. \ref{Fig: FM} (d) and (e). In the SM \cite{supplementary}, we provide a second example of $\beta$ phase with canted AFM ordering. We emphasize that $\alpha$ and $\beta$ phases share the same bulk index $\kappa=2$ and are hence topologically equivalent. Although phenomenologically distinct, an $\alpha$ phase can be transformed into a $\beta$ phase by symmetrically attaching 2d layers of Chern insulator to the side surfaces. In other words, $\alpha$ phase can be 
connected to $\beta$ phase by {\it only} closing its surface gaps, which can be achieved by manipulating its bulk magnetic ordering via ${\bf B}$ field \cite{supplementary}.

\begin{figure}[b]
	\centering
	\includegraphics[width=0.43\textwidth]{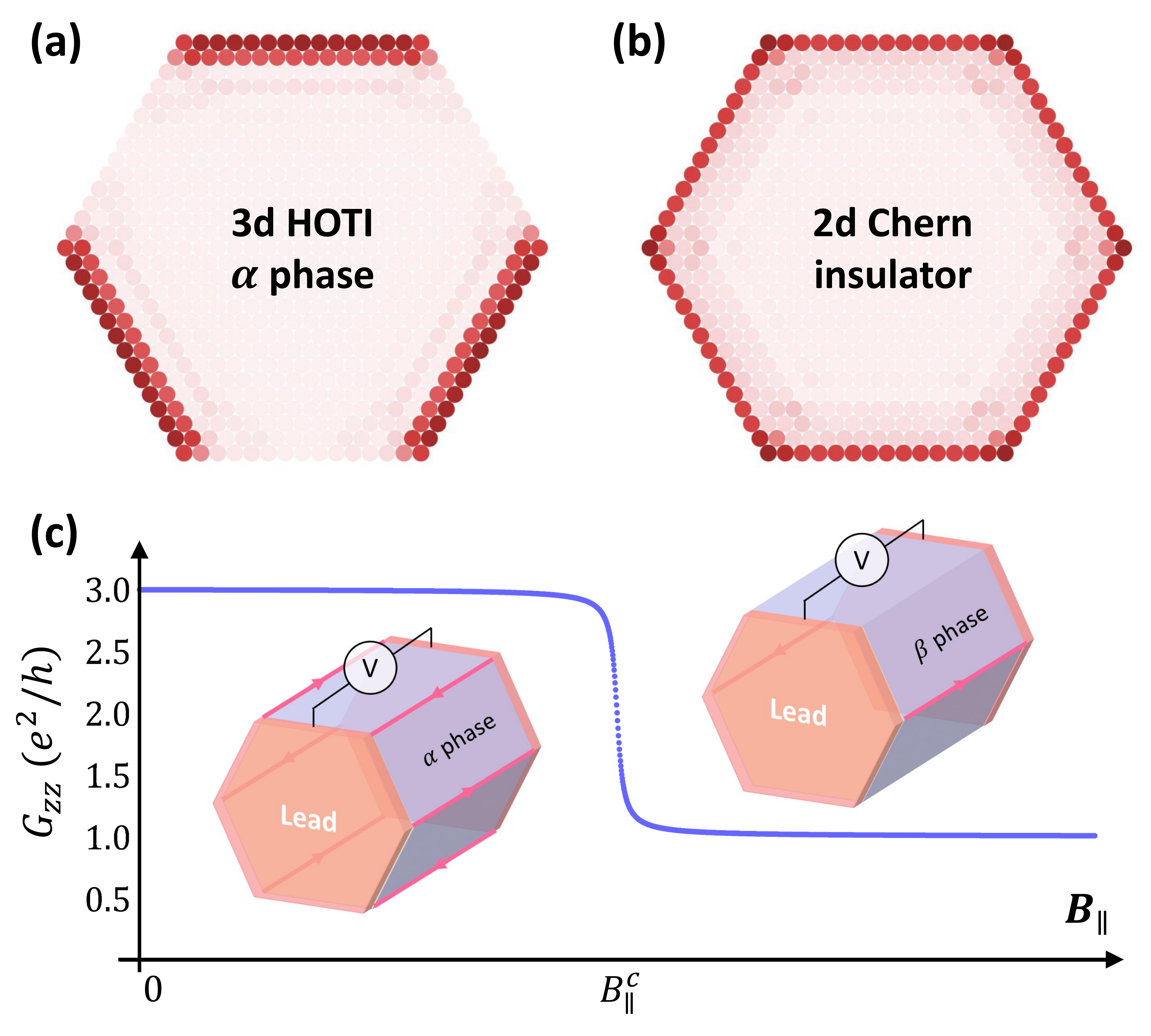}
	\caption{In-gap LDOS for (a) the top surface of a HOTI $\alpha$ phase and (b) a 2d Chern insulator. (c) The transition from HOTI $\alpha$ phase to $\beta$ phase is accompanied by a quantized jump in $G_{zz}$, which realizes a topological magnetic switch.}
	\label{Fig: STM}
\end{figure}

{\it Experiment Signatures} - For a finite-slab geometry, the chiral hinge modes in both $\alpha$ and $\beta$ phases enable in-plane quantized anomalous Hall conductance $\sigma_{xy}=e^2/h$. Notably, this conductance quantization to $e^2/h$ persists when the slab thickness grows, which signals 3d higher-order topology. In addition, we suggest using scanning tunneling microscopy to map out the in-gap local density of states (LDOS) on the top surface. As shown in Fig. \ref{Fig: STM} (a),  we predict that for a HOTI $\alpha$ phase, the LDOS on the top surface only peaks at some specific hinges or step edges. As a comparison, we also plot the in-gap LDOS peak of a 2d Chern insulator, which generally appears on all edges [Fig. \ref{Fig: STM} (b)]. This special LDOS pattern of the chiral hinge modes provides another direct experimental evidence for higher-order topology.

Meanwhile, $\alpha$ and $\beta$ phases behave distinctly in their out-of-plane transport measurements. In Fig. \ref{Fig: STM} (c), we propose measuring the two-terminal conductance $G_{zz}=I_z/V_z$ by driving a $\hat{z}$-directional current $I_z$ and measuring the voltage drop $V_z$ between top and bottom surfaces. Although quantized signals are expected for both phases, the $\alpha$ phase has $G_{zz}=3e^2/h$ while the $\beta$ phase has $G_{zz}=e^2/h$. Since an in-plane magnetic field ${\bf B}_{\parallel}$ generally drives the transition between these phases, this device realizes a ``{\it topological magnetic switch}", where the ${\bf B}_{\parallel}$ controls a quantized jump of $G_{zz}$ by a factor of 3 [Fig. \ref{Fig: STM} (c)]. Remarkably, the threshold for $B_{\parallel}$ to trigger this switching could be as low as $0.2$ T \cite{hu2019van,vidal2019topological}, which holds promise for practical topological electronics.  

{\it Conclusion} - We have proposed MnBi$_{2n}$Te$_{3n+1}$ as a possible material class for realizing higher-order M\"obius physics and various inversion-symmetric higher-order topological phases. For MnBi$_2$Te$_4$, MnBi$_4$Te$_7$, and MnBi$_6$Te$_{10}$, A-type AFM physics have been reported at zero field and signatures of canted AFM have been observed with an in-plane magnetic field \cite{gong2019experimental,hu2019van,wu2019natural,vidal2019topological,shi2019magnetic,tian2019magnetic}. Based on our topological index analysis, we expect those canted AFM systems to be exactly our proposed higher-order M\"obius insulators, when the magnetic field is carefully aligned to preserve the glide symmetry.

For both MnBi$_4$Te$_7$ and MnBi$_6$Te$_{10}$, an out-of-plane FM phase is observed at a small ($\sim$0.2 T) \cite{hu2019van,wu2019natural,vidal2019topological} or even vanishing field \cite{shi2019magnetic,tian2019magnetic}. Recent first-principles calculations suggest a non-trivial symmetry indicator $\kappa = 2$ for the FM phases in both materials \cite{vidal2019topological,tian2019magnetic}, which supports our prediction of the HOTI $\alpha$ phase in these FM systems. Moreover, the HOTI $\beta$ phase and the topological magnetic switch effect can be feasibly achieved by simply rotating the magnetic field.  

Finally, we emphasize that our theory provides a microscopic mechanism for higher-order M\"obius insulators in magnetic topological materials. For example, the higher-order M\"obius physics can also be realized on the (010) surface of the {\it afmc} phase in the axion insulator candidate EuIn$_2$As$_2$ \cite{xu2019higher} when an external in-plane field cants the magnetic moments. Similar physics can also be expected for other candidate axion insulators, such as EuSn$_2$As$_2$ \cite{li2019dirac} and EuSn$_2$P$_2$ \cite{gui2019new}. With the recent rapid developments in this field, we believe that our proposed M\"obius and higher-order topological phases should soon be experimentally realizable.    

{\it Acknowledgment} - We thank Yang-Zhi Chou, Sheng-Jie Huang, Jiabin Yu, and Zhida Song for helpful discussions. This work is supported by the Laboratory for Physical Sciences and Microsoft. R.X.Z acknowledges a JQI postdoctoral fellowship.

\bibliography{MBT}

\begin{thebibliography}{55}%
\makeatletter
\providecommand \@ifxundefined [1]{%
 \@ifx{#1\undefined}
}%
\providecommand \@ifnum [1]{%
 \ifnum #1\expandafter \@firstoftwo
 \else \expandafter \@secondoftwo
 \fi
}%
\providecommand \@ifx [1]{%
 \ifx #1\expandafter \@firstoftwo
 \else \expandafter \@secondoftwo
 \fi
}%
\providecommand \natexlab [1]{#1}%
\providecommand \enquote  [1]{``#1''}%
\providecommand \bibnamefont  [1]{#1}%
\providecommand \bibfnamefont [1]{#1}%
\providecommand \citenamefont [1]{#1}%
\providecommand \href@noop [0]{\@secondoftwo}%
\providecommand \href [0]{\begingroup \@sanitize@url \@href}%
\providecommand \@href[1]{\@@startlink{#1}\@@href}%
\providecommand \@@href[1]{\endgroup#1\@@endlink}%
\providecommand \@sanitize@url [0]{\catcode `\\12\catcode `\$12\catcode
  `\&12\catcode `\#12\catcode `\^12\catcode `\_12\catcode `\%12\relax}%
\providecommand \@@startlink[1]{}%
\providecommand \@@endlink[0]{}%
\providecommand \url  [0]{\begingroup\@sanitize@url \@url }%
\providecommand \@url [1]{\endgroup\@href {#1}{\urlprefix }}%
\providecommand \urlprefix  [0]{URL }%
\providecommand \Eprint [0]{\href }%
\providecommand \doibase [0]{http://dx.doi.org/}%
\providecommand \selectlanguage [0]{\@gobble}%
\providecommand \bibinfo  [0]{\@secondoftwo}%
\providecommand \bibfield  [0]{\@secondoftwo}%
\providecommand \translation [1]{[#1]}%
\providecommand \BibitemOpen [0]{}%
\providecommand \bibitemStop [0]{}%
\providecommand \bibitemNoStop [0]{.\EOS\space}%
\providecommand \EOS [0]{\spacefactor3000\relax}%
\providecommand \BibitemShut  [1]{\csname bibitem#1\endcsname}%
\let\auto@bib@innerbib\@empty
\bibitem [{\citenamefont {Fu}(2011)}]{fu2011topological}%
  \BibitemOpen
  \bibfield  {author} {\bibinfo {author} {\bibfnamefont {L.}~\bibnamefont
  {Fu}},\ }\href {\doibase 10.1103/PhysRevLett.106.106802} {\bibfield
  {journal} {\bibinfo  {journal} {Phys. Rev. Lett.}\ }\textbf {\bibinfo
  {volume} {106}},\ \bibinfo {pages} {106802} (\bibinfo {year}
  {2011})}\BibitemShut {NoStop}%
\bibitem [{\citenamefont {Hsieh}\ \emph {et~al.}(2012)\citenamefont {Hsieh},
  \citenamefont {Lin}, \citenamefont {Liu}, \citenamefont {Duan}, \citenamefont
  {Bansil},\ and\ \citenamefont {Fu}}]{hsieh2012topological}%
  \BibitemOpen
  \bibfield  {author} {\bibinfo {author} {\bibfnamefont {T.~H.}\ \bibnamefont
  {Hsieh}}, \bibinfo {author} {\bibfnamefont {H.}~\bibnamefont {Lin}}, \bibinfo
  {author} {\bibfnamefont {J.}~\bibnamefont {Liu}}, \bibinfo {author}
  {\bibfnamefont {W.}~\bibnamefont {Duan}}, \bibinfo {author} {\bibfnamefont
  {A.}~\bibnamefont {Bansil}}, \ and\ \bibinfo {author} {\bibfnamefont
  {L.}~\bibnamefont {Fu}},\ }\href {\doibase 10.1038/ncomms1969} {\bibfield
  {journal} {\bibinfo  {journal} {Nature Communications}\ }\textbf {\bibinfo
  {volume} {3}},\ \bibinfo {pages} {982} (\bibinfo {year} {2012})}\BibitemShut
  {NoStop}%
\bibitem [{\citenamefont {Mong}\ \emph {et~al.}(2010)\citenamefont {Mong},
  \citenamefont {Essin},\ and\ \citenamefont {Moore}}]{mong2010antiferro}%
  \BibitemOpen
  \bibfield  {author} {\bibinfo {author} {\bibfnamefont {R.~S.~K.}\
  \bibnamefont {Mong}}, \bibinfo {author} {\bibfnamefont {A.~M.}\ \bibnamefont
  {Essin}}, \ and\ \bibinfo {author} {\bibfnamefont {J.~E.}\ \bibnamefont
  {Moore}},\ }\href {\doibase 10.1103/PhysRevB.81.245209} {\bibfield  {journal}
  {\bibinfo  {journal} {Phys. Rev. B}\ }\textbf {\bibinfo {volume} {81}},\
  \bibinfo {pages} {245209} (\bibinfo {year} {2010})}\BibitemShut {NoStop}%
\bibitem [{\citenamefont {Ando}\ and\ \citenamefont
  {Fu}(2015)}]{ando2015topological}%
  \BibitemOpen
  \bibfield  {author} {\bibinfo {author} {\bibfnamefont {Y.}~\bibnamefont
  {Ando}}\ and\ \bibinfo {author} {\bibfnamefont {L.}~\bibnamefont {Fu}},\
  }\href {\doibase 10.1146/annurev-conmatphys-031214-014501} {\bibfield
  {journal} {\bibinfo  {journal} {Annual Review of Condensed Matter Physics}\
  }\textbf {\bibinfo {volume} {6}},\ \bibinfo {pages} {361} (\bibinfo {year}
  {2015})}\BibitemShut {NoStop}%
\bibitem [{\citenamefont {Liu}\ \emph {et~al.}(2014)\citenamefont {Liu},
  \citenamefont {Zhang},\ and\ \citenamefont
  {VanLeeuwen}}]{liu2014topological}%
  \BibitemOpen
  \bibfield  {author} {\bibinfo {author} {\bibfnamefont {C.-X.}\ \bibnamefont
  {Liu}}, \bibinfo {author} {\bibfnamefont {R.-X.}\ \bibnamefont {Zhang}}, \
  and\ \bibinfo {author} {\bibfnamefont {B.~K.}\ \bibnamefont {VanLeeuwen}},\
  }\href {\doibase 10.1103/PhysRevB.90.085304} {\bibfield  {journal} {\bibinfo
  {journal} {Phys. Rev. B}\ }\textbf {\bibinfo {volume} {90}},\ \bibinfo
  {pages} {085304} (\bibinfo {year} {2014})}\BibitemShut {NoStop}%
\bibitem [{\citenamefont {Zhang}\ and\ \citenamefont
  {Liu}(2015)}]{zhang2015topological}%
  \BibitemOpen
  \bibfield  {author} {\bibinfo {author} {\bibfnamefont {R.-X.}\ \bibnamefont
  {Zhang}}\ and\ \bibinfo {author} {\bibfnamefont {C.-X.}\ \bibnamefont
  {Liu}},\ }\href {\doibase 10.1103/PhysRevB.91.115317} {\bibfield  {journal}
  {\bibinfo  {journal} {Phys. Rev. B}\ }\textbf {\bibinfo {volume} {91}},\
  \bibinfo {pages} {115317} (\bibinfo {year} {2015})}\BibitemShut {NoStop}%
\bibitem [{\citenamefont {Shiozaki}\ \emph {et~al.}(2015)\citenamefont
  {Shiozaki}, \citenamefont {Sato},\ and\ \citenamefont
  {Gomi}}]{shiozaki2015z2}%
  \BibitemOpen
  \bibfield  {author} {\bibinfo {author} {\bibfnamefont {K.}~\bibnamefont
  {Shiozaki}}, \bibinfo {author} {\bibfnamefont {M.}~\bibnamefont {Sato}}, \
  and\ \bibinfo {author} {\bibfnamefont {K.}~\bibnamefont {Gomi}},\ }\href
  {\doibase 10.1103/PhysRevB.91.155120} {\bibfield  {journal} {\bibinfo
  {journal} {Phys. Rev. B}\ }\textbf {\bibinfo {volume} {91}},\ \bibinfo
  {pages} {155120} (\bibinfo {year} {2015})}\BibitemShut {NoStop}%
\bibitem [{\citenamefont {Fang}\ and\ \citenamefont {Fu}(2015)}]{fang2015new}%
  \BibitemOpen
  \bibfield  {author} {\bibinfo {author} {\bibfnamefont {C.}~\bibnamefont
  {Fang}}\ and\ \bibinfo {author} {\bibfnamefont {L.}~\bibnamefont {Fu}},\
  }\href {\doibase 10.1103/PhysRevB.91.161105} {\bibfield  {journal} {\bibinfo
  {journal} {Phys. Rev. B}\ }\textbf {\bibinfo {volume} {91}},\ \bibinfo
  {pages} {161105} (\bibinfo {year} {2015})}\BibitemShut {NoStop}%
\bibitem [{\citenamefont {Wang}\ \emph {et~al.}(2016)\citenamefont {Wang},
  \citenamefont {Alexandradinata}, \citenamefont {Cava},\ and\ \citenamefont
  {Bernevig}}]{wang2016hourglass}%
  \BibitemOpen
  \bibfield  {author} {\bibinfo {author} {\bibfnamefont {Z.}~\bibnamefont
  {Wang}}, \bibinfo {author} {\bibfnamefont {A.}~\bibnamefont
  {Alexandradinata}}, \bibinfo {author} {\bibfnamefont {R.~J.}\ \bibnamefont
  {Cava}}, \ and\ \bibinfo {author} {\bibfnamefont {B.~A.}\ \bibnamefont
  {Bernevig}},\ }\href {https://doi.org/10.1038/nature17410} {\bibfield
  {journal} {\bibinfo  {journal} {Nature}\ }\textbf {\bibinfo {volume} {532}},\
  \bibinfo {pages} {189 EP } (\bibinfo {year} {2016})}\BibitemShut {NoStop}%
\bibitem [{\citenamefont {Chang}\ \emph {et~al.}(2017)\citenamefont {Chang},
  \citenamefont {Erten},\ and\ \citenamefont {Coleman}}]{chang2017mobius}%
  \BibitemOpen
  \bibfield  {author} {\bibinfo {author} {\bibfnamefont {P.-Y.}\ \bibnamefont
  {Chang}}, \bibinfo {author} {\bibfnamefont {O.}~\bibnamefont {Erten}}, \ and\
  \bibinfo {author} {\bibfnamefont {P.}~\bibnamefont {Coleman}},\ }\href
  {https://doi.org/10.1038/nphys4092} {\bibfield  {journal} {\bibinfo
  {journal} {Nature Physics}\ }\textbf {\bibinfo {volume} {13}},\ \bibinfo
  {pages} {794 EP } (\bibinfo {year} {2017})}\BibitemShut {NoStop}%
\bibitem [{\citenamefont {Benalcazar}\ \emph
  {et~al.}(2017{\natexlab{a}})\citenamefont {Benalcazar}, \citenamefont
  {Bernevig},\ and\ \citenamefont {Hughes}}]{benalcazar2017quantized}%
  \BibitemOpen
  \bibfield  {author} {\bibinfo {author} {\bibfnamefont {W.~A.}\ \bibnamefont
  {Benalcazar}}, \bibinfo {author} {\bibfnamefont {B.~A.}\ \bibnamefont
  {Bernevig}}, \ and\ \bibinfo {author} {\bibfnamefont {T.~L.}\ \bibnamefont
  {Hughes}},\ }\href {\doibase 10.1126/science.aah6442} {\bibfield  {journal}
  {\bibinfo  {journal} {Science}\ }\textbf {\bibinfo {volume} {357}},\ \bibinfo
  {pages} {61} (\bibinfo {year} {2017}{\natexlab{a}})}\BibitemShut {NoStop}%
\bibitem [{\citenamefont {Benalcazar}\ \emph
  {et~al.}(2017{\natexlab{b}})\citenamefont {Benalcazar}, \citenamefont
  {Bernevig},\ and\ \citenamefont {Hughes}}]{benalcazar2017electric}%
  \BibitemOpen
  \bibfield  {author} {\bibinfo {author} {\bibfnamefont {W.~A.}\ \bibnamefont
  {Benalcazar}}, \bibinfo {author} {\bibfnamefont {B.~A.}\ \bibnamefont
  {Bernevig}}, \ and\ \bibinfo {author} {\bibfnamefont {T.~L.}\ \bibnamefont
  {Hughes}},\ }\href {\doibase 10.1103/PhysRevB.96.245115} {\bibfield
  {journal} {\bibinfo  {journal} {Phys. Rev. B}\ }\textbf {\bibinfo {volume}
  {96}},\ \bibinfo {pages} {245115} (\bibinfo {year}
  {2017}{\natexlab{b}})}\BibitemShut {NoStop}%
\bibitem [{\citenamefont {Schindler}\ \emph
  {et~al.}(2018{\natexlab{a}})\citenamefont {Schindler}, \citenamefont {Cook},
  \citenamefont {Vergniory}, \citenamefont {Wang}, \citenamefont {Parkin},
  \citenamefont {Bernevig},\ and\ \citenamefont
  {Neupert}}]{schindler2018higher}%
  \BibitemOpen
  \bibfield  {author} {\bibinfo {author} {\bibfnamefont {F.}~\bibnamefont
  {Schindler}}, \bibinfo {author} {\bibfnamefont {A.~M.}\ \bibnamefont {Cook}},
  \bibinfo {author} {\bibfnamefont {M.~G.}\ \bibnamefont {Vergniory}}, \bibinfo
  {author} {\bibfnamefont {Z.}~\bibnamefont {Wang}}, \bibinfo {author}
  {\bibfnamefont {S.~S.~P.}\ \bibnamefont {Parkin}}, \bibinfo {author}
  {\bibfnamefont {B.~A.}\ \bibnamefont {Bernevig}}, \ and\ \bibinfo {author}
  {\bibfnamefont {T.}~\bibnamefont {Neupert}},\ }\href {\doibase
  10.1126/sciadv.aat0346} {\bibfield  {journal} {\bibinfo  {journal} {Science
  Advances}\ }\textbf {\bibinfo {volume} {4}} (\bibinfo {year}
  {2018}{\natexlab{a}}),\ 10.1126/sciadv.aat0346}\BibitemShut {NoStop}%
\bibitem [{\citenamefont {Langbehn}\ \emph {et~al.}(2017)\citenamefont
  {Langbehn}, \citenamefont {Peng}, \citenamefont {Trifunovic}, \citenamefont
  {von Oppen},\ and\ \citenamefont {Brouwer}}]{langbehn2017reflection}%
  \BibitemOpen
  \bibfield  {author} {\bibinfo {author} {\bibfnamefont {J.}~\bibnamefont
  {Langbehn}}, \bibinfo {author} {\bibfnamefont {Y.}~\bibnamefont {Peng}},
  \bibinfo {author} {\bibfnamefont {L.}~\bibnamefont {Trifunovic}}, \bibinfo
  {author} {\bibfnamefont {F.}~\bibnamefont {von Oppen}}, \ and\ \bibinfo
  {author} {\bibfnamefont {P.~W.}\ \bibnamefont {Brouwer}},\ }\href {\doibase
  10.1103/PhysRevLett.119.246401} {\bibfield  {journal} {\bibinfo  {journal}
  {Phys. Rev. Lett.}\ }\textbf {\bibinfo {volume} {119}},\ \bibinfo {pages}
  {246401} (\bibinfo {year} {2017})}\BibitemShut {NoStop}%
\bibitem [{\citenamefont {Khalaf}(2018)}]{khalaf2018higher}%
  \BibitemOpen
  \bibfield  {author} {\bibinfo {author} {\bibfnamefont {E.}~\bibnamefont
  {Khalaf}},\ }\href {\doibase 10.1103/PhysRevB.97.205136} {\bibfield
  {journal} {\bibinfo  {journal} {Phys. Rev. B}\ }\textbf {\bibinfo {volume}
  {97}},\ \bibinfo {pages} {205136} (\bibinfo {year} {2018})}\BibitemShut
  {NoStop}%
\bibitem [{\citenamefont {Khalaf}\ \emph {et~al.}(2018)\citenamefont {Khalaf},
  \citenamefont {Po}, \citenamefont {Vishwanath},\ and\ \citenamefont
  {Watanabe}}]{khalaf2018symmetry}%
  \BibitemOpen
  \bibfield  {author} {\bibinfo {author} {\bibfnamefont {E.}~\bibnamefont
  {Khalaf}}, \bibinfo {author} {\bibfnamefont {H.~C.}\ \bibnamefont {Po}},
  \bibinfo {author} {\bibfnamefont {A.}~\bibnamefont {Vishwanath}}, \ and\
  \bibinfo {author} {\bibfnamefont {H.}~\bibnamefont {Watanabe}},\ }\href
  {\doibase 10.1103/PhysRevX.8.031070} {\bibfield  {journal} {\bibinfo
  {journal} {Phys. Rev. X}\ }\textbf {\bibinfo {volume} {8}},\ \bibinfo {pages}
  {031070} (\bibinfo {year} {2018})}\BibitemShut {NoStop}%
\bibitem [{\citenamefont {Zhang}\ \emph {et~al.}(2013)\citenamefont {Zhang},
  \citenamefont {Kane},\ and\ \citenamefont {Mele}}]{zhang2013surface}%
  \BibitemOpen
  \bibfield  {author} {\bibinfo {author} {\bibfnamefont {F.}~\bibnamefont
  {Zhang}}, \bibinfo {author} {\bibfnamefont {C.~L.}\ \bibnamefont {Kane}}, \
  and\ \bibinfo {author} {\bibfnamefont {E.~J.}\ \bibnamefont {Mele}},\ }\href
  {\doibase 10.1103/PhysRevLett.110.046404} {\bibfield  {journal} {\bibinfo
  {journal} {Phys. Rev. Lett.}\ }\textbf {\bibinfo {volume} {110}},\ \bibinfo
  {pages} {046404} (\bibinfo {year} {2013})}\BibitemShut {NoStop}%
\bibitem [{\citenamefont {Schindler}\ \emph
  {et~al.}(2018{\natexlab{b}})\citenamefont {Schindler}, \citenamefont {Wang},
  \citenamefont {Vergniory}, \citenamefont {Cook}, \citenamefont {Murani},
  \citenamefont {Sengupta}, \citenamefont {Kasumov}, \citenamefont {Deblock},
  \citenamefont {Jeon}, \citenamefont {Drozdov}, \citenamefont {Bouchiat},
  \citenamefont {Gu{\'e}ron}, \citenamefont {Yazdani}, \citenamefont
  {Bernevig},\ and\ \citenamefont {Neupert}}]{schindler2018Bi}%
  \BibitemOpen
  \bibfield  {author} {\bibinfo {author} {\bibfnamefont {F.}~\bibnamefont
  {Schindler}}, \bibinfo {author} {\bibfnamefont {Z.}~\bibnamefont {Wang}},
  \bibinfo {author} {\bibfnamefont {M.~G.}\ \bibnamefont {Vergniory}}, \bibinfo
  {author} {\bibfnamefont {A.~M.}\ \bibnamefont {Cook}}, \bibinfo {author}
  {\bibfnamefont {A.}~\bibnamefont {Murani}}, \bibinfo {author} {\bibfnamefont
  {S.}~\bibnamefont {Sengupta}}, \bibinfo {author} {\bibfnamefont {A.~Y.}\
  \bibnamefont {Kasumov}}, \bibinfo {author} {\bibfnamefont {R.}~\bibnamefont
  {Deblock}}, \bibinfo {author} {\bibfnamefont {S.}~\bibnamefont {Jeon}},
  \bibinfo {author} {\bibfnamefont {I.}~\bibnamefont {Drozdov}}, \bibinfo
  {author} {\bibfnamefont {H.}~\bibnamefont {Bouchiat}}, \bibinfo {author}
  {\bibfnamefont {S.}~\bibnamefont {Gu{\'e}ron}}, \bibinfo {author}
  {\bibfnamefont {A.}~\bibnamefont {Yazdani}}, \bibinfo {author} {\bibfnamefont
  {B.~A.}\ \bibnamefont {Bernevig}}, \ and\ \bibinfo {author} {\bibfnamefont
  {T.}~\bibnamefont {Neupert}},\ }\href {\doibase 10.1038/s41567-018-0224-7}
  {\bibfield  {journal} {\bibinfo  {journal} {Nature Physics}\ }\textbf
  {\bibinfo {volume} {14}},\ \bibinfo {pages} {918} (\bibinfo {year}
  {2018}{\natexlab{b}})}\BibitemShut {NoStop}%
\bibitem [{\citenamefont {Xu}\ \emph {et~al.}(2019{\natexlab{a}})\citenamefont
  {Xu}, \citenamefont {Song}, \citenamefont {Wang}, \citenamefont {Weng},\ and\
  \citenamefont {Dai}}]{xu2019higher}%
  \BibitemOpen
  \bibfield  {author} {\bibinfo {author} {\bibfnamefont {Y.}~\bibnamefont
  {Xu}}, \bibinfo {author} {\bibfnamefont {Z.}~\bibnamefont {Song}}, \bibinfo
  {author} {\bibfnamefont {Z.}~\bibnamefont {Wang}}, \bibinfo {author}
  {\bibfnamefont {H.}~\bibnamefont {Weng}}, \ and\ \bibinfo {author}
  {\bibfnamefont {X.}~\bibnamefont {Dai}},\ }\href {\doibase
  10.1103/PhysRevLett.122.256402} {\bibfield  {journal} {\bibinfo  {journal}
  {Phys. Rev. Lett.}\ }\textbf {\bibinfo {volume} {122}},\ \bibinfo {pages}
  {256402} (\bibinfo {year} {2019}{\natexlab{a}})}\BibitemShut {NoStop}%
\bibitem [{\citenamefont {Yue}\ \emph {et~al.}(2019)\citenamefont {Yue},
  \citenamefont {Xu}, \citenamefont {Song}, \citenamefont {Weng}, \citenamefont
  {Lu}, \citenamefont {Fang},\ and\ \citenamefont {Dai}}]{yue2019symmetry}%
  \BibitemOpen
  \bibfield  {author} {\bibinfo {author} {\bibfnamefont {C.}~\bibnamefont
  {Yue}}, \bibinfo {author} {\bibfnamefont {Y.}~\bibnamefont {Xu}}, \bibinfo
  {author} {\bibfnamefont {Z.}~\bibnamefont {Song}}, \bibinfo {author}
  {\bibfnamefont {H.}~\bibnamefont {Weng}}, \bibinfo {author} {\bibfnamefont
  {Y.-M.}\ \bibnamefont {Lu}}, \bibinfo {author} {\bibfnamefont
  {C.}~\bibnamefont {Fang}}, \ and\ \bibinfo {author} {\bibfnamefont
  {X.}~\bibnamefont {Dai}},\ }\href {\doibase 10.1038/s41567-019-0457-0}
  {\bibfield  {journal} {\bibinfo  {journal} {Nature Physics}\ }\textbf
  {\bibinfo {volume} {15}},\ \bibinfo {pages} {577} (\bibinfo {year}
  {2019})}\BibitemShut {NoStop}%
\bibitem [{\citenamefont {Wang}\ \emph {et~al.}(2018)\citenamefont {Wang},
  \citenamefont {Wieder}, \citenamefont {Li}, \citenamefont {Yan},\ and\
  \citenamefont {Bernevig}}]{wang2018higher}%
  \BibitemOpen
  \bibfield  {author} {\bibinfo {author} {\bibfnamefont {Z.}~\bibnamefont
  {Wang}}, \bibinfo {author} {\bibfnamefont {B.~J.}\ \bibnamefont {Wieder}},
  \bibinfo {author} {\bibfnamefont {J.}~\bibnamefont {Li}}, \bibinfo {author}
  {\bibfnamefont {B.}~\bibnamefont {Yan}}, \ and\ \bibinfo {author}
  {\bibfnamefont {B.~A.}\ \bibnamefont {Bernevig}},\ }\href@noop {} {\bibfield
  {journal} {\bibinfo  {journal} {arXiv preprint arXiv:1806.11116}\ } (\bibinfo
  {year} {2018})}\BibitemShut {NoStop}%
\bibitem [{\citenamefont {Lee}\ \emph {et~al.}(2019{\natexlab{a}})\citenamefont
  {Lee}, \citenamefont {Kim}, \citenamefont {Ahn},\ and\ \citenamefont
  {Yang}}]{lee2019higher}%
  \BibitemOpen
  \bibfield  {author} {\bibinfo {author} {\bibfnamefont {E.}~\bibnamefont
  {Lee}}, \bibinfo {author} {\bibfnamefont {R.}~\bibnamefont {Kim}}, \bibinfo
  {author} {\bibfnamefont {J.}~\bibnamefont {Ahn}}, \ and\ \bibinfo {author}
  {\bibfnamefont {B.-J.}\ \bibnamefont {Yang}},\ }\href
  {https://arxiv.org/abs/1904.11452} {\bibfield  {journal} {\bibinfo  {journal}
  {arXiv preprint arXiv:1904.11452}\ } (\bibinfo {year}
  {2019}{\natexlab{a}})}\BibitemShut {NoStop}%
\bibitem [{\citenamefont {Sheng}\ \emph {et~al.}(2019)\citenamefont {Sheng},
  \citenamefont {Chen}, \citenamefont {Liu}, \citenamefont {Chen},
  \citenamefont {Zhao}, \citenamefont {Yu},\ and\ \citenamefont
  {Yang}}]{sheng2019two}%
  \BibitemOpen
  \bibfield  {author} {\bibinfo {author} {\bibfnamefont {X.-L.}\ \bibnamefont
  {Sheng}}, \bibinfo {author} {\bibfnamefont {C.}~\bibnamefont {Chen}},
  \bibinfo {author} {\bibfnamefont {H.}~\bibnamefont {Liu}}, \bibinfo {author}
  {\bibfnamefont {Z.}~\bibnamefont {Chen}}, \bibinfo {author} {\bibfnamefont
  {Y.}~\bibnamefont {Zhao}}, \bibinfo {author} {\bibfnamefont {Z.-M.}\
  \bibnamefont {Yu}}, \ and\ \bibinfo {author} {\bibfnamefont {S.~A.}\
  \bibnamefont {Yang}},\ }\href {https://arxiv.org/abs/1904.09985} {\bibfield
  {journal} {\bibinfo  {journal} {arXiv preprint arXiv:1904.09985}\ } (\bibinfo
  {year} {2019})}\BibitemShut {NoStop}%
\bibitem [{\citenamefont {Zhang}\ \emph {et~al.}(2019)\citenamefont {Zhang},
  \citenamefont {Shi}, \citenamefont {Zhu}, \citenamefont {Xing}, \citenamefont
  {Zhang},\ and\ \citenamefont {Wang}}]{zhang2019MBT}%
  \BibitemOpen
  \bibfield  {author} {\bibinfo {author} {\bibfnamefont {D.}~\bibnamefont
  {Zhang}}, \bibinfo {author} {\bibfnamefont {M.}~\bibnamefont {Shi}}, \bibinfo
  {author} {\bibfnamefont {T.}~\bibnamefont {Zhu}}, \bibinfo {author}
  {\bibfnamefont {D.}~\bibnamefont {Xing}}, \bibinfo {author} {\bibfnamefont
  {H.}~\bibnamefont {Zhang}}, \ and\ \bibinfo {author} {\bibfnamefont
  {J.}~\bibnamefont {Wang}},\ }\href {\doibase 10.1103/PhysRevLett.122.206401}
  {\bibfield  {journal} {\bibinfo  {journal} {Phys. Rev. Lett.}\ }\textbf
  {\bibinfo {volume} {122}},\ \bibinfo {pages} {206401} (\bibinfo {year}
  {2019})}\BibitemShut {NoStop}%
\bibitem [{\citenamefont {Otrokov}\ \emph {et~al.}(2018)\citenamefont
  {Otrokov}, \citenamefont {Klimovskikh}, \citenamefont {Bentmann},
  \citenamefont {Zeugner}, \citenamefont {Aliev}, \citenamefont {Gass},
  \citenamefont {Wolter}, \citenamefont {Koroleva}, \citenamefont {Estyunin},
  \citenamefont {Shikin} \emph {et~al.}}]{otrokov2018prediction}%
  \BibitemOpen
  \bibfield  {author} {\bibinfo {author} {\bibfnamefont {M.~M.}\ \bibnamefont
  {Otrokov}}, \bibinfo {author} {\bibfnamefont {I.~I.}\ \bibnamefont
  {Klimovskikh}}, \bibinfo {author} {\bibfnamefont {H.}~\bibnamefont
  {Bentmann}}, \bibinfo {author} {\bibfnamefont {A.}~\bibnamefont {Zeugner}},
  \bibinfo {author} {\bibfnamefont {Z.~S.}\ \bibnamefont {Aliev}}, \bibinfo
  {author} {\bibfnamefont {S.}~\bibnamefont {Gass}}, \bibinfo {author}
  {\bibfnamefont {A.~U.}\ \bibnamefont {Wolter}}, \bibinfo {author}
  {\bibfnamefont {A.~V.}\ \bibnamefont {Koroleva}}, \bibinfo {author}
  {\bibfnamefont {D.}~\bibnamefont {Estyunin}}, \bibinfo {author}
  {\bibfnamefont {A.~M.}\ \bibnamefont {Shikin}},  \emph {et~al.},\ }\href@noop
  {} {\  (\bibinfo {year} {2018})},\ \Eprint {http://arxiv.org/abs/1809.07389}
  {arXiv:1809.07389 [cond-mat.mtrl-sci]} \BibitemShut {NoStop}%
\bibitem [{\citenamefont {Gong}\ \emph {et~al.}(2019)\citenamefont {Gong},
  \citenamefont {Guo}, \citenamefont {Li}, \citenamefont {Zhu}, \citenamefont
  {Liao}, \citenamefont {Liu}, \citenamefont {Zhang}, \citenamefont {Gu},
  \citenamefont {Tang}, \citenamefont {Feng}, \citenamefont {Zhang},
  \citenamefont {Li}, \citenamefont {Song}, \citenamefont {Wang}, \citenamefont
  {Yu}, \citenamefont {Chen}, \citenamefont {Wang}, \citenamefont {Yao},
  \citenamefont {Duan}, \citenamefont {Xu}, \citenamefont {Zhang},
  \citenamefont {Ma}, \citenamefont {Xue},\ and\ \citenamefont
  {He}}]{gong2019experimental}%
  \BibitemOpen
  \bibfield  {author} {\bibinfo {author} {\bibfnamefont {Y.}~\bibnamefont
  {Gong}}, \bibinfo {author} {\bibfnamefont {J.}~\bibnamefont {Guo}}, \bibinfo
  {author} {\bibfnamefont {J.}~\bibnamefont {Li}}, \bibinfo {author}
  {\bibfnamefont {K.}~\bibnamefont {Zhu}}, \bibinfo {author} {\bibfnamefont
  {M.}~\bibnamefont {Liao}}, \bibinfo {author} {\bibfnamefont {X.}~\bibnamefont
  {Liu}}, \bibinfo {author} {\bibfnamefont {Q.}~\bibnamefont {Zhang}}, \bibinfo
  {author} {\bibfnamefont {L.}~\bibnamefont {Gu}}, \bibinfo {author}
  {\bibfnamefont {L.}~\bibnamefont {Tang}}, \bibinfo {author} {\bibfnamefont
  {X.}~\bibnamefont {Feng}}, \bibinfo {author} {\bibfnamefont {D.}~\bibnamefont
  {Zhang}}, \bibinfo {author} {\bibfnamefont {W.}~\bibnamefont {Li}}, \bibinfo
  {author} {\bibfnamefont {C.}~\bibnamefont {Song}}, \bibinfo {author}
  {\bibfnamefont {L.}~\bibnamefont {Wang}}, \bibinfo {author} {\bibfnamefont
  {P.}~\bibnamefont {Yu}}, \bibinfo {author} {\bibfnamefont {X.}~\bibnamefont
  {Chen}}, \bibinfo {author} {\bibfnamefont {Y.}~\bibnamefont {Wang}}, \bibinfo
  {author} {\bibfnamefont {H.}~\bibnamefont {Yao}}, \bibinfo {author}
  {\bibfnamefont {W.}~\bibnamefont {Duan}}, \bibinfo {author} {\bibfnamefont
  {Y.}~\bibnamefont {Xu}}, \bibinfo {author} {\bibfnamefont {S.-C.}\
  \bibnamefont {Zhang}}, \bibinfo {author} {\bibfnamefont {X.}~\bibnamefont
  {Ma}}, \bibinfo {author} {\bibfnamefont {Q.-K.}\ \bibnamefont {Xue}}, \ and\
  \bibinfo {author} {\bibfnamefont {K.}~\bibnamefont {He}},\ }\href {\doibase
  10.1088/0256-307x/36/7/076801} {\bibfield  {journal} {\bibinfo  {journal}
  {Chinese Physics Letters}\ }\textbf {\bibinfo {volume} {36}},\ \bibinfo
  {pages} {076801} (\bibinfo {year} {2019})}\BibitemShut {NoStop}%
\bibitem [{\citenamefont {Li}\ \emph {et~al.}(2019{\natexlab{a}})\citenamefont
  {Li}, \citenamefont {Li}, \citenamefont {Du}, \citenamefont {Wang},
  \citenamefont {Gu}, \citenamefont {Zhang}, \citenamefont {He}, \citenamefont
  {Duan},\ and\ \citenamefont {Xu}}]{li2019intrinsic}%
  \BibitemOpen
  \bibfield  {author} {\bibinfo {author} {\bibfnamefont {J.}~\bibnamefont
  {Li}}, \bibinfo {author} {\bibfnamefont {Y.}~\bibnamefont {Li}}, \bibinfo
  {author} {\bibfnamefont {S.}~\bibnamefont {Du}}, \bibinfo {author}
  {\bibfnamefont {Z.}~\bibnamefont {Wang}}, \bibinfo {author} {\bibfnamefont
  {B.-L.}\ \bibnamefont {Gu}}, \bibinfo {author} {\bibfnamefont {S.-C.}\
  \bibnamefont {Zhang}}, \bibinfo {author} {\bibfnamefont {K.}~\bibnamefont
  {He}}, \bibinfo {author} {\bibfnamefont {W.}~\bibnamefont {Duan}}, \ and\
  \bibinfo {author} {\bibfnamefont {Y.}~\bibnamefont {Xu}},\ }\href {\doibase
  10.1126/sciadv.aaw5685} {\bibfield  {journal} {\bibinfo  {journal} {Science
  Advances}\ }\textbf {\bibinfo {volume} {5}} (\bibinfo {year}
  {2019}{\natexlab{a}}),\ 10.1126/sciadv.aaw5685}\BibitemShut {NoStop}%
\bibitem [{\citenamefont {Vidal}\ \emph
  {et~al.}(2019{\natexlab{a}})\citenamefont {Vidal}, \citenamefont {Bentmann},
  \citenamefont {Peixoto}, \citenamefont {Zeugner}, \citenamefont {Moser},
  \citenamefont {Min}, \citenamefont {Schatz}, \citenamefont {Ki\ss{}ner},
  \citenamefont {\"Unzelmann}, \citenamefont {Fornari}, \citenamefont {Vasili},
  \citenamefont {Valvidares}, \citenamefont {Sakamoto}, \citenamefont {Mondal},
  \citenamefont {Fujii}, \citenamefont {Vobornik}, \citenamefont {Jung},
  \citenamefont {Cacho}, \citenamefont {Kim}, \citenamefont {Koch},
  \citenamefont {Jozwiak}, \citenamefont {Bostwick}, \citenamefont {Denlinger},
  \citenamefont {Rotenberg}, \citenamefont {Buck}, \citenamefont {Hoesch},
  \citenamefont {Diekmann}, \citenamefont {Rohlf}, \citenamefont {Kall\"ane},
  \citenamefont {Rossnagel}, \citenamefont {Otrokov}, \citenamefont {Chulkov},
  \citenamefont {Ruck}, \citenamefont {Isaeva},\ and\ \citenamefont
  {Reinert}}]{vidal2019surface}%
  \BibitemOpen
  \bibfield  {author} {\bibinfo {author} {\bibfnamefont {R.~C.}\ \bibnamefont
  {Vidal}}, \bibinfo {author} {\bibfnamefont {H.}~\bibnamefont {Bentmann}},
  \bibinfo {author} {\bibfnamefont {T.~R.~F.}\ \bibnamefont {Peixoto}},
  \bibinfo {author} {\bibfnamefont {A.}~\bibnamefont {Zeugner}}, \bibinfo
  {author} {\bibfnamefont {S.}~\bibnamefont {Moser}}, \bibinfo {author}
  {\bibfnamefont {C.-H.}\ \bibnamefont {Min}}, \bibinfo {author} {\bibfnamefont
  {S.}~\bibnamefont {Schatz}}, \bibinfo {author} {\bibfnamefont
  {K.}~\bibnamefont {Ki\ss{}ner}}, \bibinfo {author} {\bibfnamefont
  {M.}~\bibnamefont {\"Unzelmann}}, \bibinfo {author} {\bibfnamefont {C.~I.}\
  \bibnamefont {Fornari}}, \bibinfo {author} {\bibfnamefont {H.~B.}\
  \bibnamefont {Vasili}}, \bibinfo {author} {\bibfnamefont {M.}~\bibnamefont
  {Valvidares}}, \bibinfo {author} {\bibfnamefont {K.}~\bibnamefont
  {Sakamoto}}, \bibinfo {author} {\bibfnamefont {D.}~\bibnamefont {Mondal}},
  \bibinfo {author} {\bibfnamefont {J.}~\bibnamefont {Fujii}}, \bibinfo
  {author} {\bibfnamefont {I.}~\bibnamefont {Vobornik}}, \bibinfo {author}
  {\bibfnamefont {S.}~\bibnamefont {Jung}}, \bibinfo {author} {\bibfnamefont
  {C.}~\bibnamefont {Cacho}}, \bibinfo {author} {\bibfnamefont {T.~K.}\
  \bibnamefont {Kim}}, \bibinfo {author} {\bibfnamefont {R.~J.}\ \bibnamefont
  {Koch}}, \bibinfo {author} {\bibfnamefont {C.}~\bibnamefont {Jozwiak}},
  \bibinfo {author} {\bibfnamefont {A.}~\bibnamefont {Bostwick}}, \bibinfo
  {author} {\bibfnamefont {J.~D.}\ \bibnamefont {Denlinger}}, \bibinfo {author}
  {\bibfnamefont {E.}~\bibnamefont {Rotenberg}}, \bibinfo {author}
  {\bibfnamefont {J.}~\bibnamefont {Buck}}, \bibinfo {author} {\bibfnamefont
  {M.}~\bibnamefont {Hoesch}}, \bibinfo {author} {\bibfnamefont
  {F.}~\bibnamefont {Diekmann}}, \bibinfo {author} {\bibfnamefont
  {S.}~\bibnamefont {Rohlf}}, \bibinfo {author} {\bibfnamefont
  {M.}~\bibnamefont {Kall\"ane}}, \bibinfo {author} {\bibfnamefont
  {K.}~\bibnamefont {Rossnagel}}, \bibinfo {author} {\bibfnamefont {M.~M.}\
  \bibnamefont {Otrokov}}, \bibinfo {author} {\bibfnamefont {E.~V.}\
  \bibnamefont {Chulkov}}, \bibinfo {author} {\bibfnamefont {M.}~\bibnamefont
  {Ruck}}, \bibinfo {author} {\bibfnamefont {A.}~\bibnamefont {Isaeva}}, \ and\
  \bibinfo {author} {\bibfnamefont {F.}~\bibnamefont {Reinert}},\ }\href
  {\doibase 10.1103/PhysRevB.100.121104} {\bibfield  {journal} {\bibinfo
  {journal} {Phys. Rev. B}\ }\textbf {\bibinfo {volume} {100}},\ \bibinfo
  {pages} {121104} (\bibinfo {year} {2019}{\natexlab{a}})}\BibitemShut
  {NoStop}%
\bibitem [{\citenamefont {Hu}\ \emph {et~al.}(2019{\natexlab{a}})\citenamefont
  {Hu}, \citenamefont {Zhou}, \citenamefont {Liu}, \citenamefont {Liu},
  \citenamefont {Hao}, \citenamefont {Emmanouilidou}, \citenamefont {Sun},
  \citenamefont {Liu}, \citenamefont {Brawer}, \citenamefont {Ramirez},
  \citenamefont {Cao}, \citenamefont {Liu}, \citenamefont {Dessau},\ and\
  \citenamefont {Ni}}]{hu2019van}%
  \BibitemOpen
  \bibfield  {author} {\bibinfo {author} {\bibfnamefont {C.}~\bibnamefont
  {Hu}}, \bibinfo {author} {\bibfnamefont {X.}~\bibnamefont {Zhou}}, \bibinfo
  {author} {\bibfnamefont {P.}~\bibnamefont {Liu}}, \bibinfo {author}
  {\bibfnamefont {J.}~\bibnamefont {Liu}}, \bibinfo {author} {\bibfnamefont
  {P.}~\bibnamefont {Hao}}, \bibinfo {author} {\bibfnamefont {E.}~\bibnamefont
  {Emmanouilidou}}, \bibinfo {author} {\bibfnamefont {H.}~\bibnamefont {Sun}},
  \bibinfo {author} {\bibfnamefont {Y.}~\bibnamefont {Liu}}, \bibinfo {author}
  {\bibfnamefont {H.}~\bibnamefont {Brawer}}, \bibinfo {author} {\bibfnamefont
  {A.~P.}\ \bibnamefont {Ramirez}}, \bibinfo {author} {\bibfnamefont
  {H.}~\bibnamefont {Cao}}, \bibinfo {author} {\bibfnamefont {Q.}~\bibnamefont
  {Liu}}, \bibinfo {author} {\bibfnamefont {D.}~\bibnamefont {Dessau}}, \ and\
  \bibinfo {author} {\bibfnamefont {N.}~\bibnamefont {Ni}},\ }\href@noop {} {\
  (\bibinfo {year} {2019}{\natexlab{a}})},\ \Eprint
  {http://arxiv.org/abs/1905.02154} {arXiv:1905.02154 [cond-mat.mtrl-sci]}
  \BibitemShut {NoStop}%
\bibitem [{\citenamefont {Wu}\ \emph {et~al.}(2019)\citenamefont {Wu},
  \citenamefont {Liu}, \citenamefont {Sasase}, \citenamefont {Ienaga},
  \citenamefont {Obata}, \citenamefont {Yukawa}, \citenamefont {Horiba},
  \citenamefont {Kumigashira}, \citenamefont {Okuma}, \citenamefont
  {Inoshita},\ and\ \citenamefont {Hosono}}]{wu2019natural}%
  \BibitemOpen
  \bibfield  {author} {\bibinfo {author} {\bibfnamefont {J.}~\bibnamefont
  {Wu}}, \bibinfo {author} {\bibfnamefont {F.}~\bibnamefont {Liu}}, \bibinfo
  {author} {\bibfnamefont {M.}~\bibnamefont {Sasase}}, \bibinfo {author}
  {\bibfnamefont {K.}~\bibnamefont {Ienaga}}, \bibinfo {author} {\bibfnamefont
  {Y.}~\bibnamefont {Obata}}, \bibinfo {author} {\bibfnamefont
  {R.}~\bibnamefont {Yukawa}}, \bibinfo {author} {\bibfnamefont
  {K.}~\bibnamefont {Horiba}}, \bibinfo {author} {\bibfnamefont
  {H.}~\bibnamefont {Kumigashira}}, \bibinfo {author} {\bibfnamefont
  {S.}~\bibnamefont {Okuma}}, \bibinfo {author} {\bibfnamefont
  {T.}~\bibnamefont {Inoshita}}, \ and\ \bibinfo {author} {\bibfnamefont
  {H.}~\bibnamefont {Hosono}},\ }\href@noop {} {\  (\bibinfo {year} {2019})},\
  \Eprint {http://arxiv.org/abs/1905.02385} {arXiv:1905.02385
  [cond-mat.mtrl-sci]} \BibitemShut {NoStop}%
\bibitem [{\citenamefont {Li}\ \emph {et~al.}(2019{\natexlab{b}})\citenamefont
  {Li}, \citenamefont {Gao}, \citenamefont {Duan}, \citenamefont {Xu},
  \citenamefont {Zhu}, \citenamefont {Tian}, \citenamefont {Fan}, \citenamefont
  {Rao}, \citenamefont {Huang}, \citenamefont {Li} \emph
  {et~al.}}]{li2019dirac}%
  \BibitemOpen
  \bibfield  {author} {\bibinfo {author} {\bibfnamefont {H.}~\bibnamefont
  {Li}}, \bibinfo {author} {\bibfnamefont {S.-Y.}\ \bibnamefont {Gao}},
  \bibinfo {author} {\bibfnamefont {S.-F.}\ \bibnamefont {Duan}}, \bibinfo
  {author} {\bibfnamefont {Y.-F.}\ \bibnamefont {Xu}}, \bibinfo {author}
  {\bibfnamefont {K.-J.}\ \bibnamefont {Zhu}}, \bibinfo {author} {\bibfnamefont
  {S.-J.}\ \bibnamefont {Tian}}, \bibinfo {author} {\bibfnamefont {W.-H.}\
  \bibnamefont {Fan}}, \bibinfo {author} {\bibfnamefont {Z.-C.}\ \bibnamefont
  {Rao}}, \bibinfo {author} {\bibfnamefont {J.-R.}\ \bibnamefont {Huang}},
  \bibinfo {author} {\bibfnamefont {J.-J.}\ \bibnamefont {Li}},  \emph
  {et~al.},\ }\href@noop {} {\  (\bibinfo {year} {2019}{\natexlab{b}})},\
  \Eprint {http://arxiv.org/abs/1907.06491} {arXiv:1907.06491
  [cond-mat.mtrl-sci]} \BibitemShut {NoStop}%
\bibitem [{\citenamefont {Hao}\ \emph {et~al.}(2019)\citenamefont {Hao},
  \citenamefont {Liu}, \citenamefont {Feng}, \citenamefont {Ma}, \citenamefont
  {Schwier}, \citenamefont {Arita}, \citenamefont {Kumar}, \citenamefont {Hu},
  \citenamefont {Lu}, \citenamefont {Zeng} \emph {et~al.}}]{hao2019gapless}%
  \BibitemOpen
  \bibfield  {author} {\bibinfo {author} {\bibfnamefont {Y.-J.}\ \bibnamefont
  {Hao}}, \bibinfo {author} {\bibfnamefont {P.}~\bibnamefont {Liu}}, \bibinfo
  {author} {\bibfnamefont {Y.}~\bibnamefont {Feng}}, \bibinfo {author}
  {\bibfnamefont {X.-M.}\ \bibnamefont {Ma}}, \bibinfo {author} {\bibfnamefont
  {E.~F.}\ \bibnamefont {Schwier}}, \bibinfo {author} {\bibfnamefont
  {M.}~\bibnamefont {Arita}}, \bibinfo {author} {\bibfnamefont
  {S.}~\bibnamefont {Kumar}}, \bibinfo {author} {\bibfnamefont
  {C.}~\bibnamefont {Hu}}, \bibinfo {author} {\bibfnamefont {R.}~\bibnamefont
  {Lu}}, \bibinfo {author} {\bibfnamefont {M.}~\bibnamefont {Zeng}},  \emph
  {et~al.},\ }\href@noop {} {\  (\bibinfo {year} {2019})},\ \Eprint
  {http://arxiv.org/abs/1907.03722} {arXiv:1907.03722 [cond-mat.mtrl-sci]}
  \BibitemShut {NoStop}%
\bibitem [{\citenamefont {Chen}\ \emph {et~al.}(2019)\citenamefont {Chen},
  \citenamefont {Xu}, \citenamefont {Li}, \citenamefont {Li}, \citenamefont
  {Zhang}, \citenamefont {Li}, \citenamefont {Wu}, \citenamefont {Liang},
  \citenamefont {Chen}, \citenamefont {Jung} \emph
  {et~al.}}]{chen2019topological}%
  \BibitemOpen
  \bibfield  {author} {\bibinfo {author} {\bibfnamefont {Y.}~\bibnamefont
  {Chen}}, \bibinfo {author} {\bibfnamefont {L.}~\bibnamefont {Xu}}, \bibinfo
  {author} {\bibfnamefont {J.}~\bibnamefont {Li}}, \bibinfo {author}
  {\bibfnamefont {Y.}~\bibnamefont {Li}}, \bibinfo {author} {\bibfnamefont
  {C.}~\bibnamefont {Zhang}}, \bibinfo {author} {\bibfnamefont
  {H.}~\bibnamefont {Li}}, \bibinfo {author} {\bibfnamefont {Y.}~\bibnamefont
  {Wu}}, \bibinfo {author} {\bibfnamefont {A.}~\bibnamefont {Liang}}, \bibinfo
  {author} {\bibfnamefont {C.}~\bibnamefont {Chen}}, \bibinfo {author}
  {\bibfnamefont {S.}~\bibnamefont {Jung}},  \emph {et~al.},\ }\href@noop {} {\
   (\bibinfo {year} {2019})},\ \Eprint {http://arxiv.org/abs/1907.05119}
  {arXiv:1907.05119 [cond-mat.mtrl-sci]} \BibitemShut {NoStop}%
\bibitem [{\citenamefont {Vidal}\ \emph
  {et~al.}(2019{\natexlab{b}})\citenamefont {Vidal}, \citenamefont {Zeugner},
  \citenamefont {Facio}, \citenamefont {Ray}, \citenamefont {Haghighi},
  \citenamefont {Wolter}, \citenamefont {Bohorquez}, \citenamefont {Caglieris},
  \citenamefont {Moser}, \citenamefont {Figgemeier} \emph
  {et~al.}}]{vidal2019topological}%
  \BibitemOpen
  \bibfield  {author} {\bibinfo {author} {\bibfnamefont {R.~C.}\ \bibnamefont
  {Vidal}}, \bibinfo {author} {\bibfnamefont {A.}~\bibnamefont {Zeugner}},
  \bibinfo {author} {\bibfnamefont {J.~I.}\ \bibnamefont {Facio}}, \bibinfo
  {author} {\bibfnamefont {R.}~\bibnamefont {Ray}}, \bibinfo {author}
  {\bibfnamefont {M.~H.}\ \bibnamefont {Haghighi}}, \bibinfo {author}
  {\bibfnamefont {A.~U.}\ \bibnamefont {Wolter}}, \bibinfo {author}
  {\bibfnamefont {L.~T.~C.}\ \bibnamefont {Bohorquez}}, \bibinfo {author}
  {\bibfnamefont {F.}~\bibnamefont {Caglieris}}, \bibinfo {author}
  {\bibfnamefont {S.}~\bibnamefont {Moser}}, \bibinfo {author} {\bibfnamefont
  {T.}~\bibnamefont {Figgemeier}},  \emph {et~al.},\ }\href@noop {} {\
  (\bibinfo {year} {2019}{\natexlab{b}})},\ \Eprint
  {http://arxiv.org/abs/1906.08394} {arXiv:1906.08394 [cond-mat.mtrl-sci]}
  \BibitemShut {NoStop}%
\bibitem [{\citenamefont {Shi}\ \emph {et~al.}(2019)\citenamefont {Shi},
  \citenamefont {Lei}, \citenamefont {Zhu}, \citenamefont {Ma}, \citenamefont
  {Cui}, \citenamefont {Sun}, \citenamefont {Ying},\ and\ \citenamefont
  {Chen}}]{shi2019magnetic}%
  \BibitemOpen
  \bibfield  {author} {\bibinfo {author} {\bibfnamefont {M.~Z.}\ \bibnamefont
  {Shi}}, \bibinfo {author} {\bibfnamefont {B.}~\bibnamefont {Lei}}, \bibinfo
  {author} {\bibfnamefont {C.~S.}\ \bibnamefont {Zhu}}, \bibinfo {author}
  {\bibfnamefont {D.~H.}\ \bibnamefont {Ma}}, \bibinfo {author} {\bibfnamefont
  {J.~H.}\ \bibnamefont {Cui}}, \bibinfo {author} {\bibfnamefont {Z.~L.}\
  \bibnamefont {Sun}}, \bibinfo {author} {\bibfnamefont {J.~J.}\ \bibnamefont
  {Ying}}, \ and\ \bibinfo {author} {\bibfnamefont {X.~H.}\ \bibnamefont
  {Chen}},\ }\href@noop {} {\  (\bibinfo {year} {2019})},\ \Eprint
  {http://arxiv.org/abs/1910.08912} {arXiv:1910.08912 [cond-mat.mtrl-sci]}
  \BibitemShut {NoStop}%
\bibitem [{\citenamefont {Ding}\ \emph {et~al.}(2019)\citenamefont {Ding},
  \citenamefont {Hu}, \citenamefont {Ye}, \citenamefont {Feng}, \citenamefont
  {Ni},\ and\ \citenamefont {Cao}}]{ding2019crystal}%
  \BibitemOpen
  \bibfield  {author} {\bibinfo {author} {\bibfnamefont {L.}~\bibnamefont
  {Ding}}, \bibinfo {author} {\bibfnamefont {C.}~\bibnamefont {Hu}}, \bibinfo
  {author} {\bibfnamefont {F.}~\bibnamefont {Ye}}, \bibinfo {author}
  {\bibfnamefont {E.}~\bibnamefont {Feng}}, \bibinfo {author} {\bibfnamefont
  {N.}~\bibnamefont {Ni}}, \ and\ \bibinfo {author} {\bibfnamefont
  {H.}~\bibnamefont {Cao}},\ }\href@noop {} {\  (\bibinfo {year} {2019})},\
  \Eprint {http://arxiv.org/abs/1910.06248} {arXiv:1910.06248
  [cond-mat.str-el]} \BibitemShut {NoStop}%
\bibitem [{\citenamefont {Tian}\ \emph {et~al.}(2019)\citenamefont {Tian},
  \citenamefont {Gao}, \citenamefont {Nie}, \citenamefont {Qian}, \citenamefont
  {Gong}, \citenamefont {Fu}, \citenamefont {Li}, \citenamefont {Fan},
  \citenamefont {Zhang}, \citenamefont {Kondo}, \citenamefont {Shin},
  \citenamefont {Adell}, \citenamefont {Fedderwitz}, \citenamefont {Ding},
  \citenamefont {Wang}, \citenamefont {Qian},\ and\ \citenamefont
  {Lei}}]{tian2019magnetic}%
  \BibitemOpen
  \bibfield  {author} {\bibinfo {author} {\bibfnamefont {S.}~\bibnamefont
  {Tian}}, \bibinfo {author} {\bibfnamefont {S.}~\bibnamefont {Gao}}, \bibinfo
  {author} {\bibfnamefont {S.}~\bibnamefont {Nie}}, \bibinfo {author}
  {\bibfnamefont {Y.}~\bibnamefont {Qian}}, \bibinfo {author} {\bibfnamefont
  {C.}~\bibnamefont {Gong}}, \bibinfo {author} {\bibfnamefont {Y.}~\bibnamefont
  {Fu}}, \bibinfo {author} {\bibfnamefont {H.}~\bibnamefont {Li}}, \bibinfo
  {author} {\bibfnamefont {W.}~\bibnamefont {Fan}}, \bibinfo {author}
  {\bibfnamefont {P.}~\bibnamefont {Zhang}}, \bibinfo {author} {\bibfnamefont
  {T.}~\bibnamefont {Kondo}}, \bibinfo {author} {\bibfnamefont
  {S.}~\bibnamefont {Shin}}, \bibinfo {author} {\bibfnamefont {J.}~\bibnamefont
  {Adell}}, \bibinfo {author} {\bibfnamefont {H.}~\bibnamefont {Fedderwitz}},
  \bibinfo {author} {\bibfnamefont {H.}~\bibnamefont {Ding}}, \bibinfo {author}
  {\bibfnamefont {Z.}~\bibnamefont {Wang}}, \bibinfo {author} {\bibfnamefont
  {T.}~\bibnamefont {Qian}}, \ and\ \bibinfo {author} {\bibfnamefont
  {H.}~\bibnamefont {Lei}},\ }\href@noop {} {\  (\bibinfo {year} {2019})},\
  \Eprint {http://arxiv.org/abs/1910.10101} {arXiv:1910.10101
  [cond-mat.mtrl-sci]} \BibitemShut {NoStop}%
\bibitem [{\citenamefont {Xu}\ \emph {et~al.}(2019{\natexlab{b}})\citenamefont
  {Xu}, \citenamefont {Mao}, \citenamefont {Wang}, \citenamefont {Li},
  \citenamefont {Chen}, \citenamefont {Xia}, \citenamefont {Li}, \citenamefont
  {Zhang}, \citenamefont {Zheng}, \citenamefont {Huang}, \citenamefont {Zhang},
  \citenamefont {Cui}, \citenamefont {Liang}, \citenamefont {Xia},
  \citenamefont {Su}, \citenamefont {Jung}, \citenamefont {Cacho},
  \citenamefont {Wang}, \citenamefont {Li}, \citenamefont {Xu}, \citenamefont
  {Guo}, \citenamefont {Yang}, \citenamefont {Liu},\ and\ \citenamefont
  {Chen}}]{xu2019persistent}%
  \BibitemOpen
  \bibfield  {author} {\bibinfo {author} {\bibfnamefont {L.~X.}\ \bibnamefont
  {Xu}}, \bibinfo {author} {\bibfnamefont {Y.~H.}\ \bibnamefont {Mao}},
  \bibinfo {author} {\bibfnamefont {H.~Y.}\ \bibnamefont {Wang}}, \bibinfo
  {author} {\bibfnamefont {J.~H.}\ \bibnamefont {Li}}, \bibinfo {author}
  {\bibfnamefont {Y.~J.}\ \bibnamefont {Chen}}, \bibinfo {author}
  {\bibfnamefont {Y.~Y.~Y.}\ \bibnamefont {Xia}}, \bibinfo {author}
  {\bibfnamefont {Y.~W.}\ \bibnamefont {Li}}, \bibinfo {author} {\bibfnamefont
  {J.}~\bibnamefont {Zhang}}, \bibinfo {author} {\bibfnamefont {H.~J.}\
  \bibnamefont {Zheng}}, \bibinfo {author} {\bibfnamefont {K.}~\bibnamefont
  {Huang}}, \bibinfo {author} {\bibfnamefont {C.~F.}\ \bibnamefont {Zhang}},
  \bibinfo {author} {\bibfnamefont {S.~T.}\ \bibnamefont {Cui}}, \bibinfo
  {author} {\bibfnamefont {A.~J.}\ \bibnamefont {Liang}}, \bibinfo {author}
  {\bibfnamefont {W.}~\bibnamefont {Xia}}, \bibinfo {author} {\bibfnamefont
  {H.}~\bibnamefont {Su}}, \bibinfo {author} {\bibfnamefont {S.~W.}\
  \bibnamefont {Jung}}, \bibinfo {author} {\bibfnamefont {C.}~\bibnamefont
  {Cacho}}, \bibinfo {author} {\bibfnamefont {M.~X.}\ \bibnamefont {Wang}},
  \bibinfo {author} {\bibfnamefont {G.}~\bibnamefont {Li}}, \bibinfo {author}
  {\bibfnamefont {Y.}~\bibnamefont {Xu}}, \bibinfo {author} {\bibfnamefont
  {Y.~F.}\ \bibnamefont {Guo}}, \bibinfo {author} {\bibfnamefont {L.~X.}\
  \bibnamefont {Yang}}, \bibinfo {author} {\bibfnamefont {Z.~K.}\ \bibnamefont
  {Liu}}, \ and\ \bibinfo {author} {\bibfnamefont {Y.~L.}\ \bibnamefont
  {Chen}},\ }\href@noop {} {\  (\bibinfo {year} {2019}{\natexlab{b}})},\
  \Eprint {http://arxiv.org/abs/1910.11014} {arXiv:1910.11014
  [cond-mat.mtrl-sci]} \BibitemShut {NoStop}%
\bibitem [{\citenamefont {Hu}\ \emph {et~al.}(2019{\natexlab{b}})\citenamefont
  {Hu}, \citenamefont {Xu}, \citenamefont {Shi}, \citenamefont {Luo},
  \citenamefont {Peng}, \citenamefont {Wang}, \citenamefont {Ying},
  \citenamefont {Wu}, \citenamefont {Liu}, \citenamefont {Zhang}, \citenamefont
  {Chen}, \citenamefont {Xu}, \citenamefont {Chen},\ and\ \citenamefont
  {He}}]{hu2019universal}%
  \BibitemOpen
  \bibfield  {author} {\bibinfo {author} {\bibfnamefont {Y.}~\bibnamefont
  {Hu}}, \bibinfo {author} {\bibfnamefont {L.}~\bibnamefont {Xu}}, \bibinfo
  {author} {\bibfnamefont {M.}~\bibnamefont {Shi}}, \bibinfo {author}
  {\bibfnamefont {A.}~\bibnamefont {Luo}}, \bibinfo {author} {\bibfnamefont
  {S.}~\bibnamefont {Peng}}, \bibinfo {author} {\bibfnamefont {Z.~Y.}\
  \bibnamefont {Wang}}, \bibinfo {author} {\bibfnamefont {J.~J.}\ \bibnamefont
  {Ying}}, \bibinfo {author} {\bibfnamefont {T.}~\bibnamefont {Wu}}, \bibinfo
  {author} {\bibfnamefont {Z.~K.}\ \bibnamefont {Liu}}, \bibinfo {author}
  {\bibfnamefont {C.~F.}\ \bibnamefont {Zhang}}, \bibinfo {author}
  {\bibfnamefont {Y.~L.}\ \bibnamefont {Chen}}, \bibinfo {author}
  {\bibfnamefont {G.}~\bibnamefont {Xu}}, \bibinfo {author} {\bibfnamefont
  {X.~H.}\ \bibnamefont {Chen}}, \ and\ \bibinfo {author} {\bibfnamefont
  {J.~F.}\ \bibnamefont {He}},\ }\href@noop {} {\  (\bibinfo {year}
  {2019}{\natexlab{b}})},\ \Eprint {http://arxiv.org/abs/1910.11323}
  {arXiv:1910.11323 [cond-mat.mtrl-sci]} \BibitemShut {NoStop}%
\bibitem [{\citenamefont {Lee}\ \emph {et~al.}(2019{\natexlab{b}})\citenamefont
  {Lee}, \citenamefont {Zhu}, \citenamefont {Wang}, \citenamefont {Miao},
  \citenamefont {Pillsbury}, \citenamefont {Yi}, \citenamefont {Kempinger},
  \citenamefont {Hu}, \citenamefont {Heikes}, \citenamefont {Quarterman},
  \citenamefont {Ratcliff}, \citenamefont {Borchers}, \citenamefont {Zhang},
  \citenamefont {Ke}, \citenamefont {Graf}, \citenamefont {Alem}, \citenamefont
  {Chang}, \citenamefont {Samarth},\ and\ \citenamefont {Mao}}]{lee2019spin}%
  \BibitemOpen
  \bibfield  {author} {\bibinfo {author} {\bibfnamefont {S.~H.}\ \bibnamefont
  {Lee}}, \bibinfo {author} {\bibfnamefont {Y.}~\bibnamefont {Zhu}}, \bibinfo
  {author} {\bibfnamefont {Y.}~\bibnamefont {Wang}}, \bibinfo {author}
  {\bibfnamefont {L.}~\bibnamefont {Miao}}, \bibinfo {author} {\bibfnamefont
  {T.}~\bibnamefont {Pillsbury}}, \bibinfo {author} {\bibfnamefont
  {H.}~\bibnamefont {Yi}}, \bibinfo {author} {\bibfnamefont {S.}~\bibnamefont
  {Kempinger}}, \bibinfo {author} {\bibfnamefont {J.}~\bibnamefont {Hu}},
  \bibinfo {author} {\bibfnamefont {C.~A.}\ \bibnamefont {Heikes}}, \bibinfo
  {author} {\bibfnamefont {P.}~\bibnamefont {Quarterman}}, \bibinfo {author}
  {\bibfnamefont {W.}~\bibnamefont {Ratcliff}}, \bibinfo {author}
  {\bibfnamefont {J.~A.}\ \bibnamefont {Borchers}}, \bibinfo {author}
  {\bibfnamefont {H.}~\bibnamefont {Zhang}}, \bibinfo {author} {\bibfnamefont
  {X.}~\bibnamefont {Ke}}, \bibinfo {author} {\bibfnamefont {D.}~\bibnamefont
  {Graf}}, \bibinfo {author} {\bibfnamefont {N.}~\bibnamefont {Alem}}, \bibinfo
  {author} {\bibfnamefont {C.-Z.}\ \bibnamefont {Chang}}, \bibinfo {author}
  {\bibfnamefont {N.}~\bibnamefont {Samarth}}, \ and\ \bibinfo {author}
  {\bibfnamefont {Z.}~\bibnamefont {Mao}},\ }\href {\doibase
  10.1103/PhysRevResearch.1.012011} {\bibfield  {journal} {\bibinfo  {journal}
  {Phys. Rev. Research}\ }\textbf {\bibinfo {volume} {1}},\ \bibinfo {pages}
  {012011} (\bibinfo {year} {2019}{\natexlab{b}})}\BibitemShut {NoStop}%
\bibitem [{\citenamefont {Yan}\ \emph {et~al.}(2019{\natexlab{a}})\citenamefont
  {Yan}, \citenamefont {Liu}, \citenamefont {Parker}, \citenamefont {McGuire},\
  and\ \citenamefont {Sales}}]{yan2019atype}%
  \BibitemOpen
  \bibfield  {author} {\bibinfo {author} {\bibfnamefont {J.~Q.}\ \bibnamefont
  {Yan}}, \bibinfo {author} {\bibfnamefont {Y.~H.}\ \bibnamefont {Liu}},
  \bibinfo {author} {\bibfnamefont {D.}~\bibnamefont {Parker}}, \bibinfo
  {author} {\bibfnamefont {M.~A.}\ \bibnamefont {McGuire}}, \ and\ \bibinfo
  {author} {\bibfnamefont {B.~C.}\ \bibnamefont {Sales}},\ }\href@noop {} {\
  (\bibinfo {year} {2019}{\natexlab{a}})},\ \Eprint
  {http://arxiv.org/abs/1910.06273} {arXiv:1910.06273 [cond-mat.mtrl-sci]}
  \BibitemShut {NoStop}%
\bibitem [{\citenamefont {Yan}\ \emph {et~al.}(2019{\natexlab{b}})\citenamefont
  {Yan}, \citenamefont {Zhang}, \citenamefont {Heitmann}, \citenamefont
  {Huang}, \citenamefont {Chen}, \citenamefont {Cheng}, \citenamefont {Wu},
  \citenamefont {Vaknin}, \citenamefont {Sales},\ and\ \citenamefont
  {McQueeney}}]{yan2019crystal}%
  \BibitemOpen
  \bibfield  {author} {\bibinfo {author} {\bibfnamefont {J.-Q.}\ \bibnamefont
  {Yan}}, \bibinfo {author} {\bibfnamefont {Q.}~\bibnamefont {Zhang}}, \bibinfo
  {author} {\bibfnamefont {T.}~\bibnamefont {Heitmann}}, \bibinfo {author}
  {\bibfnamefont {Z.}~\bibnamefont {Huang}}, \bibinfo {author} {\bibfnamefont
  {K.~Y.}\ \bibnamefont {Chen}}, \bibinfo {author} {\bibfnamefont {J.-G.}\
  \bibnamefont {Cheng}}, \bibinfo {author} {\bibfnamefont {W.}~\bibnamefont
  {Wu}}, \bibinfo {author} {\bibfnamefont {D.}~\bibnamefont {Vaknin}}, \bibinfo
  {author} {\bibfnamefont {B.~C.}\ \bibnamefont {Sales}}, \ and\ \bibinfo
  {author} {\bibfnamefont {R.~J.}\ \bibnamefont {McQueeney}},\ }\href {\doibase
  10.1103/PhysRevMaterials.3.064202} {\bibfield  {journal} {\bibinfo  {journal}
  {Phys. Rev. Materials}\ }\textbf {\bibinfo {volume} {3}},\ \bibinfo {pages}
  {064202} (\bibinfo {year} {2019}{\natexlab{b}})}\BibitemShut {NoStop}%
\bibitem [{\citenamefont {Deng}\ \emph {et~al.}(2019)\citenamefont {Deng},
  \citenamefont {Yu}, \citenamefont {Shi}, \citenamefont {Wang}, \citenamefont
  {Chen},\ and\ \citenamefont {Zhang}}]{deng2019magnetic}%
  \BibitemOpen
  \bibfield  {author} {\bibinfo {author} {\bibfnamefont {Y.}~\bibnamefont
  {Deng}}, \bibinfo {author} {\bibfnamefont {Y.}~\bibnamefont {Yu}}, \bibinfo
  {author} {\bibfnamefont {M.~Z.}\ \bibnamefont {Shi}}, \bibinfo {author}
  {\bibfnamefont {J.}~\bibnamefont {Wang}}, \bibinfo {author} {\bibfnamefont
  {X.~H.}\ \bibnamefont {Chen}}, \ and\ \bibinfo {author} {\bibfnamefont
  {Y.}~\bibnamefont {Zhang}},\ }\href@noop {} {\  (\bibinfo {year} {2019})},\
  \Eprint {http://arxiv.org/abs/1904.11468} {arXiv:1904.11468
  [cond-mat.mtrl-sci]} \BibitemShut {NoStop}%
\bibitem [{\citenamefont {Ge}\ \emph {et~al.}(2019)\citenamefont {Ge},
  \citenamefont {Liu}, \citenamefont {Li}, \citenamefont {Li}, \citenamefont
  {Luo}, \citenamefont {Wu}, \citenamefont {Xu},\ and\ \citenamefont
  {Wang}}]{ge2019highchernnumber}%
  \BibitemOpen
  \bibfield  {author} {\bibinfo {author} {\bibfnamefont {J.}~\bibnamefont
  {Ge}}, \bibinfo {author} {\bibfnamefont {Y.}~\bibnamefont {Liu}}, \bibinfo
  {author} {\bibfnamefont {J.}~\bibnamefont {Li}}, \bibinfo {author}
  {\bibfnamefont {H.}~\bibnamefont {Li}}, \bibinfo {author} {\bibfnamefont
  {T.}~\bibnamefont {Luo}}, \bibinfo {author} {\bibfnamefont {Y.}~\bibnamefont
  {Wu}}, \bibinfo {author} {\bibfnamefont {Y.}~\bibnamefont {Xu}}, \ and\
  \bibinfo {author} {\bibfnamefont {J.}~\bibnamefont {Wang}},\ }\href@noop {}
  {\  (\bibinfo {year} {2019})},\ \Eprint {http://arxiv.org/abs/1907.09947}
  {arXiv:1907.09947 [cond-mat.mes-hall]} \BibitemShut {NoStop}%
\bibitem [{\citenamefont {Liu}\ \emph {et~al.}(2019)\citenamefont {Liu},
  \citenamefont {Wang}, \citenamefont {Li}, \citenamefont {Wu}, \citenamefont
  {Li}, \citenamefont {Li}, \citenamefont {He}, \citenamefont {Xu},
  \citenamefont {Zhang},\ and\ \citenamefont {Wang}}]{liu2019quantum}%
  \BibitemOpen
  \bibfield  {author} {\bibinfo {author} {\bibfnamefont {C.}~\bibnamefont
  {Liu}}, \bibinfo {author} {\bibfnamefont {Y.}~\bibnamefont {Wang}}, \bibinfo
  {author} {\bibfnamefont {H.}~\bibnamefont {Li}}, \bibinfo {author}
  {\bibfnamefont {Y.}~\bibnamefont {Wu}}, \bibinfo {author} {\bibfnamefont
  {Y.}~\bibnamefont {Li}}, \bibinfo {author} {\bibfnamefont {J.}~\bibnamefont
  {Li}}, \bibinfo {author} {\bibfnamefont {K.}~\bibnamefont {He}}, \bibinfo
  {author} {\bibfnamefont {Y.}~\bibnamefont {Xu}}, \bibinfo {author}
  {\bibfnamefont {J.}~\bibnamefont {Zhang}}, \ and\ \bibinfo {author}
  {\bibfnamefont {Y.}~\bibnamefont {Wang}},\ }\href@noop {} {\  (\bibinfo
  {year} {2019})},\ \Eprint {http://arxiv.org/abs/1905.00715} {arXiv:1905.00715
  [cond-mat.mes-hall]} \BibitemShut {NoStop}%
\bibitem [{foo()}]{footnote1}%
  \BibitemOpen
  \href@noop {} {}\bibinfo {note} {Our lattice model generally describes the
  topological physics in both MnBi$_{2n}$Te$_{3n+1}$ compounds with $n=2,5$
  (space group $P\bar{3}m1$) and those with $n=1,3,4$ (space group $R\bar{3}m$)
  \cite{ding2019crystal}. In the latter case, the $\hat{z}$ axis in our lattice
  model should be viewed as the $c$-axis in the rhombohedral
  lattice}\BibitemShut {NoStop}%
\bibitem [{\citenamefont {Liu}\ \emph {et~al.}(2010)\citenamefont {Liu},
  \citenamefont {Qi}, \citenamefont {Zhang}, \citenamefont {Dai}, \citenamefont
  {Fang},\ and\ \citenamefont {Zhang}}]{liu2010model}%
  \BibitemOpen
  \bibfield  {author} {\bibinfo {author} {\bibfnamefont {C.-X.}\ \bibnamefont
  {Liu}}, \bibinfo {author} {\bibfnamefont {X.-L.}\ \bibnamefont {Qi}},
  \bibinfo {author} {\bibfnamefont {H.}~\bibnamefont {Zhang}}, \bibinfo
  {author} {\bibfnamefont {X.}~\bibnamefont {Dai}}, \bibinfo {author}
  {\bibfnamefont {Z.}~\bibnamefont {Fang}}, \ and\ \bibinfo {author}
  {\bibfnamefont {S.-C.}\ \bibnamefont {Zhang}},\ }\href {\doibase
  10.1103/PhysRevB.82.045122} {\bibfield  {journal} {\bibinfo  {journal} {Phys.
  Rev. B}\ }\textbf {\bibinfo {volume} {82}},\ \bibinfo {pages} {045122}
  (\bibinfo {year} {2010})}\BibitemShut {NoStop}%
\bibitem [{sup()}]{supplementary}%
  \BibitemOpen
  \href@noop {} {}\bibinfo {note} {See Supplemental Material at XX for the full
  lattice model, equivalent relations among different symmetry indicators,
  Wilson loop characterization for M\"obius fermions, an effective surface
  theory for different HOTI phases, an example of HOTI $\beta$ phase with
  canted AFM, and a theory for field-induced canted AFM}\BibitemShut {NoStop}%
\bibitem [{\citenamefont {Wieder}\ \emph {et~al.}(2018)\citenamefont {Wieder},
  \citenamefont {Bradlyn}, \citenamefont {Wang}, \citenamefont {Cano},
  \citenamefont {Kim}, \citenamefont {Kim}, \citenamefont {Rappe},
  \citenamefont {Kane},\ and\ \citenamefont {Bernevig}}]{wieder2018wallpaper}%
  \BibitemOpen
  \bibfield  {author} {\bibinfo {author} {\bibfnamefont {B.~J.}\ \bibnamefont
  {Wieder}}, \bibinfo {author} {\bibfnamefont {B.}~\bibnamefont {Bradlyn}},
  \bibinfo {author} {\bibfnamefont {Z.}~\bibnamefont {Wang}}, \bibinfo {author}
  {\bibfnamefont {J.}~\bibnamefont {Cano}}, \bibinfo {author} {\bibfnamefont
  {Y.}~\bibnamefont {Kim}}, \bibinfo {author} {\bibfnamefont {H.-S.~D.}\
  \bibnamefont {Kim}}, \bibinfo {author} {\bibfnamefont {A.~M.}\ \bibnamefont
  {Rappe}}, \bibinfo {author} {\bibfnamefont {C.~L.}\ \bibnamefont {Kane}}, \
  and\ \bibinfo {author} {\bibfnamefont {B.~A.}\ \bibnamefont {Bernevig}},\
  }\href {\doibase 10.1126/science.aan2802} {\bibfield  {journal} {\bibinfo
  {journal} {Science}\ }\textbf {\bibinfo {volume} {361}},\ \bibinfo {pages}
  {246} (\bibinfo {year} {2018})}\BibitemShut {NoStop}%
\bibitem [{\citenamefont {Turner}\ \emph {et~al.}(2012)\citenamefont {Turner},
  \citenamefont {Zhang}, \citenamefont {Mong},\ and\ \citenamefont
  {Vishwanath}}]{turner2012quantized}%
  \BibitemOpen
  \bibfield  {author} {\bibinfo {author} {\bibfnamefont {A.~M.}\ \bibnamefont
  {Turner}}, \bibinfo {author} {\bibfnamefont {Y.}~\bibnamefont {Zhang}},
  \bibinfo {author} {\bibfnamefont {R.~S.~K.}\ \bibnamefont {Mong}}, \ and\
  \bibinfo {author} {\bibfnamefont {A.}~\bibnamefont {Vishwanath}},\ }\href
  {\doibase 10.1103/PhysRevB.85.165120} {\bibfield  {journal} {\bibinfo
  {journal} {Phys. Rev. B}\ }\textbf {\bibinfo {volume} {85}},\ \bibinfo
  {pages} {165120} (\bibinfo {year} {2012})}\BibitemShut {NoStop}%
\bibitem [{\citenamefont {Ono}\ and\ \citenamefont
  {Watanabe}(2018)}]{ono2018unified}%
  \BibitemOpen
  \bibfield  {author} {\bibinfo {author} {\bibfnamefont {S.}~\bibnamefont
  {Ono}}\ and\ \bibinfo {author} {\bibfnamefont {H.}~\bibnamefont {Watanabe}},\
  }\href {\doibase 10.1103/PhysRevB.98.115150} {\bibfield  {journal} {\bibinfo
  {journal} {Phys. Rev. B}\ }\textbf {\bibinfo {volume} {98}},\ \bibinfo
  {pages} {115150} (\bibinfo {year} {2018})}\BibitemShut {NoStop}%
\bibitem [{\citenamefont {Wieder}\ and\ \citenamefont
  {Bernevig}(2018)}]{wieder2018axion}%
  \BibitemOpen
  \bibfield  {author} {\bibinfo {author} {\bibfnamefont {B.~J.}\ \bibnamefont
  {Wieder}}\ and\ \bibinfo {author} {\bibfnamefont {B.~A.}\ \bibnamefont
  {Bernevig}},\ }\href@noop {} {\  (\bibinfo {year} {2018})},\ \Eprint
  {http://arxiv.org/abs/1810.02373} {arXiv:1810.02373 [cond-mat.mes-hall]}
  \BibitemShut {NoStop}%
\bibitem [{\citenamefont {Kim}\ \emph {et~al.}(2019)\citenamefont {Kim},
  \citenamefont {Shiozaki},\ and\ \citenamefont {Murakami}}]{kim2019glide}%
  \BibitemOpen
  \bibfield  {author} {\bibinfo {author} {\bibfnamefont {H.}~\bibnamefont
  {Kim}}, \bibinfo {author} {\bibfnamefont {K.}~\bibnamefont {Shiozaki}}, \
  and\ \bibinfo {author} {\bibfnamefont {S.}~\bibnamefont {Murakami}},\ }\href
  {\doibase 10.1103/PhysRevB.100.165202} {\bibfield  {journal} {\bibinfo
  {journal} {Phys. Rev. B}\ }\textbf {\bibinfo {volume} {100}},\ \bibinfo
  {pages} {165202} (\bibinfo {year} {2019})}\BibitemShut {NoStop}%
\bibitem [{\citenamefont {Gui}\ \emph {et~al.}(2019)\citenamefont {Gui},
  \citenamefont {Pletikosic}, \citenamefont {Cao}, \citenamefont {Tien},
  \citenamefont {Xu}, \citenamefont {Zhong}, \citenamefont {Wang},
  \citenamefont {Chang}, \citenamefont {Jia}, \citenamefont {Valla},
  \citenamefont {Xie},\ and\ \citenamefont {Cava}}]{gui2019new}%
  \BibitemOpen
  \bibfield  {author} {\bibinfo {author} {\bibfnamefont {X.}~\bibnamefont
  {Gui}}, \bibinfo {author} {\bibfnamefont {I.}~\bibnamefont {Pletikosic}},
  \bibinfo {author} {\bibfnamefont {H.}~\bibnamefont {Cao}}, \bibinfo {author}
  {\bibfnamefont {H.-J.}\ \bibnamefont {Tien}}, \bibinfo {author}
  {\bibfnamefont {X.}~\bibnamefont {Xu}}, \bibinfo {author} {\bibfnamefont
  {R.}~\bibnamefont {Zhong}}, \bibinfo {author} {\bibfnamefont
  {G.}~\bibnamefont {Wang}}, \bibinfo {author} {\bibfnamefont {T.-R.}\
  \bibnamefont {Chang}}, \bibinfo {author} {\bibfnamefont {S.}~\bibnamefont
  {Jia}}, \bibinfo {author} {\bibfnamefont {T.}~\bibnamefont {Valla}}, \bibinfo
  {author} {\bibfnamefont {W.}~\bibnamefont {Xie}}, \ and\ \bibinfo {author}
  {\bibfnamefont {R.~J.}\ \bibnamefont {Cava}},\ }\href@noop {} {\  (\bibinfo
  {year} {2019})},\ \Eprint {http://arxiv.org/abs/1903.03888} {arXiv:1903.03888
  [cond-mat.mtrl-sci]} \BibitemShut {NoStop}%
\bibitem [{\citenamefont {Baltz}\ \emph {et~al.}(2018)\citenamefont {Baltz},
  \citenamefont {Manchon}, \citenamefont {Tsoi}, \citenamefont {Moriyama},
  \citenamefont {Ono},\ and\ \citenamefont {Tserkovnyak}}]{baltz2018antiferro}%
  \BibitemOpen
  \bibfield  {author} {\bibinfo {author} {\bibfnamefont {V.}~\bibnamefont
  {Baltz}}, \bibinfo {author} {\bibfnamefont {A.}~\bibnamefont {Manchon}},
  \bibinfo {author} {\bibfnamefont {M.}~\bibnamefont {Tsoi}}, \bibinfo {author}
  {\bibfnamefont {T.}~\bibnamefont {Moriyama}}, \bibinfo {author}
  {\bibfnamefont {T.}~\bibnamefont {Ono}}, \ and\ \bibinfo {author}
  {\bibfnamefont {Y.}~\bibnamefont {Tserkovnyak}},\ }\href {\doibase
  10.1103/RevModPhys.90.015005} {\bibfield  {journal} {\bibinfo  {journal}
  {Rev. Mod. Phys.}\ }\textbf {\bibinfo {volume} {90}},\ \bibinfo {pages}
  {015005} (\bibinfo {year} {2018})}\BibitemShut {NoStop}%
\end{thebibliography}%

\onecolumngrid

\subsection{\large Supplemental Material for ``M\"{o}bius Insulator and Higher-Order Topology in MnBi$_{2n}$Te$_{3n+1}$"}

\section{Appendix A: Lattice Model}
We start by defining our effective lattice model. The continuum model in the main text is regularized on a 3d hexagonal lattice, which is defined by the lattice constants ${\bf a}_1 = (1,0,0)^T,\ {\bf a}_2 = \frac{1}{2}(1,\sqrt{3},0)^T,\ {\bf a}_3=(0,0,1)^T$. For our purpose, we further define the crystal momenta $(k_1,k_2,k_z)$ as
\begin{eqnarray}
k_1 = k_x,\ k_2 = \frac{1}{2}(k_x + \sqrt{3} k_y). 
\end{eqnarray}
The full lattice model is given by
\begin{eqnarray}
H = H_\text{lat} + H_{ex}
\end{eqnarray}
where the exchange-coupling term $H_{ex}$ is defined in Eq. [1] in the main text. The normal part 
\bea
H_\text{lat} = \begin{pmatrix}
	h & h_{AB} \\
	h_{AB}^{\dagger}  & h \\
\end{pmatrix},
\eea  
Here the intra-layer Hamiltonian is
\begin{eqnarray}
h &=& (\tilde{C} - \frac{4}{3} C_2 [\cos k_1 + \cos k_2 + \cos (k_1 - k_2)]) \mathbb{I}_4 + \frac{2v}{3} (\sin k_1 + \frac{1}{2} [\sin (k_1-k_2) + \sin k_2]) \Gamma_1 \nonumber \\
&& + (\frac{v}{\sqrt{3}} [\sin k_2 - \sin(k_1 - k_2)]) \Gamma_2 + w [-\sin k_1 + \sin k_2 + \sin (k_1-k_2)] \Gamma_4 \nonumber \\
&& + (\tilde{M}-\frac{4}{3}M_2 [\cos k_1 + \cos k_2 + \cos (k_1 - k_2)])\Gamma_5,
\end{eqnarray}
where $\tilde{C} = C_0 + 2C_1 + 4 C_2$ and $\tilde{M} = M_0 + 2M_1 + 4 M_2$. The hopping between layer A and layer B is described by 
\begin{eqnarray}
h_{AB} = -2C_1 \cos\frac{k_z}{2} \mathbb{I}_4 + 2v_z\sin \frac{k_z}{2} \Gamma_3 - 2M_1 \cos\frac{k_z}{2} \Gamma_5.
\end{eqnarray}
By expanding $H_\text{lat}$ around $\Gamma$ point, it reproduces the continuum Hamiltonian in the main text.

We have used $M_0 = -2, M_1 = 1, M_2 = 1, C_0 = 0, C_1 = 0.3, C_2 = 0.4, v_z = 1, v = 0.5, m = 0.4, w = 1.5$ as our choice of parameters throughout the work.  

\section{Appendix B: Equivalent Relations Among Symmetry Indicators}

In this section, we discuss the equivalent relations among the inversion symmetry indicators for various topological phases that appear in our system. In a time-reversal-symmetric system with inversion symmetry, we can define a $\mathbb{Z}_4$ symmetry indicator
\begin{equation}
\kappa_\Theta = \frac{1}{4} \sum_{k_i} (n_+ - n_-)\ \ (\text{mod }4).
\end{equation}
Here $n_+$ and $n_-$ are the numbers of occupied bands with $+$ and $-$ parity eigenvalues, respectively, which are summed over eight time-reversal-invariant momenta $k_i$ in a 3d BZ. While an odd (even) $\kappa$ implies existence (absence) of a strong TI phase, $\kappa=2$ characterizes a HOTI phase with 1d helical hinge states.

When AFM is introduced and the TRS is broken to an AFM TRS, the band topology is characterized by a similar $\mathbb{Z}_2$ inversion indicator $\xi\in\mathbb{Z}_2$ with
\begin{equation}
\xi = \frac{1}{4} \sum_{k_i} (n_+ - n_-)\ \ (\text{mod }2).
\label{Eq: xi}
\end{equation}
Clearly, by definition, 
\bea
\xi = \kappa_\Theta\ \ (\text{mod }2).
\eea
This leads to the following statement:
\begin{itemize}
	\item By adiabatically breaking TRS to an AFM TRS while preserving inversion, a 3d strong TI with $\kappa_\Theta=1,3$ is connected to an AFM TI with $\xi=1$.
\end{itemize} 

For the magnetic higher-order topological physics discussed in our system, it is indicated by a new $\mathbb{Z}_4$ indicator 
\begin{equation}
\kappa = \frac{1}{2} \sum_{k_i} (n_+ - n_-)\ \ (\text{mod }4),
\end{equation}
which is Eq. [2] in the main text. To see the connection between $\xi$ and $\kappa$, we note that the parity eigenvalues obey 
\bea
\sum_{k_i} (n_+ - n_-) = 4 (2l+\xi),
\eea
following Eq. \ref{Eq: xi}, where $l\in \mathbb{Z}$ is an integer. Therefore, 
\bea
\kappa &=&  4m+2\xi\ \ (\text{mod }4) \nonumber \\
&\equiv& 2\xi\ \ (\text{mod }4).
\eea
As a result, we arrive at an important conclusion:
\begin{itemize}
	\item By adiabatically breaking $\Theta_M$ while preserving inversion, a 3d AFM TI with $\xi=1$ is connected to an axion insulator with $\kappa=2$ and higher-order topology.
\end{itemize} 
Here we have included the BZ folding effect of AFM in the definitions of $\kappa,\kappa_\Theta$, and $\xi$. 

\section{Appendix C: Wilson Loop Characterization for M\"obius Fermion}

\begin{figure}[t]
	\centering
	\includegraphics[width=0.85\textwidth]{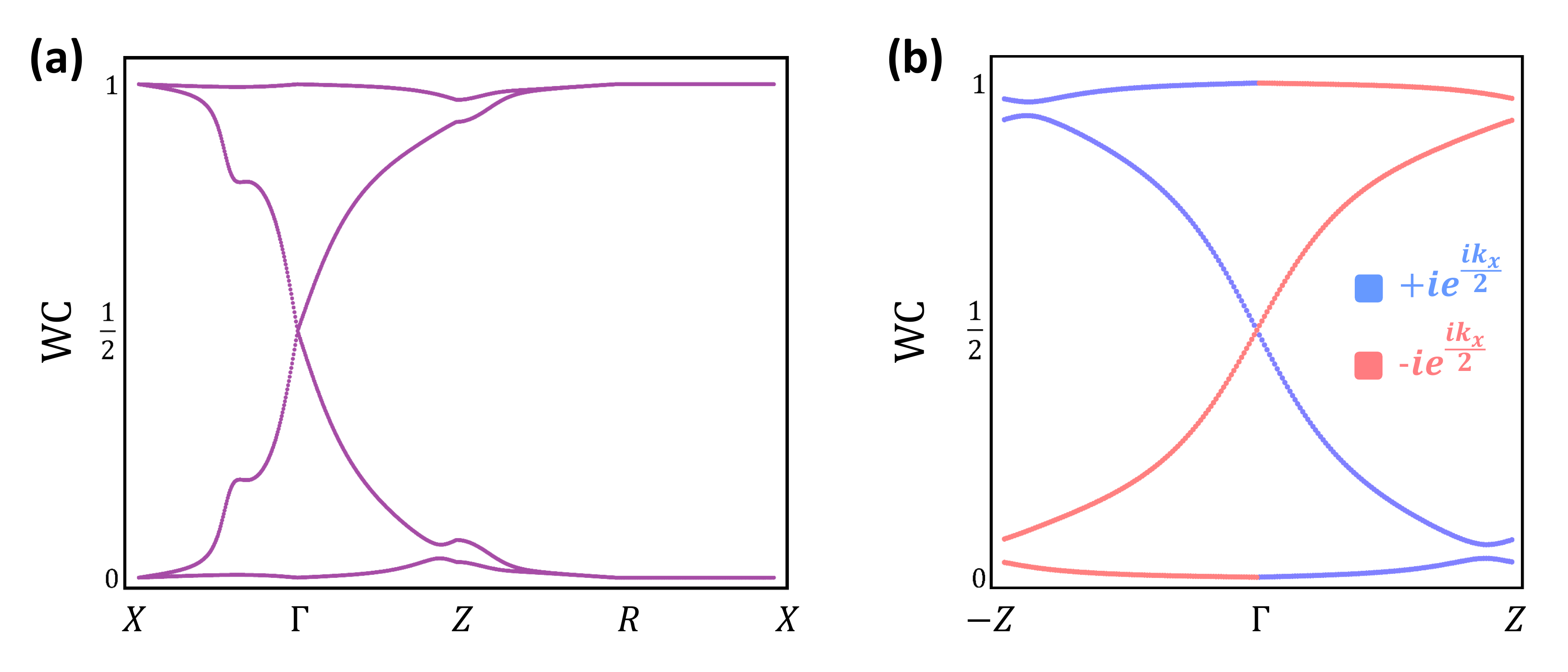}
	\caption{(a) Gapless WC flows along the high symmetry lines reveals nontrivial surface state on the (010) surface. (b) The M\"obius pattern of WC flows along the glide-invariant line. Different colors denote different glide-mirror sectors.}
	\label{Fig: Wilson Loop}
\end{figure}

In this section, we discuss the nontrivial Wilson loop spectrum that characterizes the topological nature of the higher-order M\"obius insulator. The Wilson loop along a closed path $L$ in the momentum space is defined as 
\bea
{\cal W}_L = \text{exp}[i\oint_L {\bf A} d{\bf k}],
\eea
where the non-Abelian Berry connection
\bea
{\bf A}_{mn}({\bf k}) = i\langle u_{m}({\bf k}) | \nabla_{\bf k} | u_{n}({\bf k}) \rangle.
\eea
For given $k_x$ and $k_z$, the phases of the eigenvalues for the Wilson loop 
\bea
{\cal W}_y (k_x,k_z) = \text{exp}[i\int_{0}^{2\pi} d k_y{\bf A} (k_y) ]
\eea
are the Wannier centers (WC), or equivalently the 1d electronic polarization along $\hat{y}$. Physically, the evolution of WCs in the 2d BZ spanned by $k_x$ and $k_z$ describe the charge pumping process on the $(010)$ surface. 

In Fig. \ref{Fig: Wilson Loop} (a), we calculate the evolution of WC flows along the high symmetry lines that are perpendicular to the $k_y$ axis. Crucially, the two WC flows show a nontrivial gapless winding pattern along the high symmetry lines, featuring a topological gapless surface state on the $(010)$ surface. This WC pattern is similar to that of the hourglass fermion in Ref. \cite{wang2016hourglass}.

As shown in Fig. \ref{Fig: Wilson Loop} (b), we plot the WC flows along the glide-invariant line (GIL) from $-Z$ to $Z$ via $\Gamma$. Since the Hamiltonian is block-diagonal along GIL, the WC flow for each glide-mirror sector can be individually plotted. In particular, we find that the WC flow along GIL also has a M\"obius nature, manifesting the bulk-boundary correspondence.

\section{Appendix D: Effective Surface Theory}

In this section, we establish an effective surface theory to understand the formation of chiral hinge modes and the surface topological transition of HOTI $\alpha$ and $\beta$ phases. To start with, consider a Hamiltonian $H({\bf k})$ for HOTI $\alpha$ phase, which describes a 3d topological insulator with out-of-plane FM. 
\begin{eqnarray}
H ({\bf k})= \begin{pmatrix}
e({\bf k})+M({\bf k})+g_z & v_z k_z & 0 & v(k_x-ik_y) \\
v_zk_z & e({\bf k})-M({\bf k})+g_z & v(k_x-ik_y) & 0 \\
0 & v(k_x+ik_y) & e({\bf k})+M({\bf k})-g_z & -v_zk_z \\
v(k_x+ik_y) & 0 & -v_zk_z & e({\bf k})-M({\bf k})-g_z \\
\end{pmatrix},
\end{eqnarray}
with $M({\bf k})=M_0 + M_1 k_z^2 + M_2 (k_x^2+k_y^2)$ and $C({\bf k})= C_2 k_y^2$. Here we have ignored the $k_x^2$ and $k_z^2$ terms in $C{\bf k}$ for simplicity. $g_z$ chracterizes the FM exchange coupling parallel to $\hat{z}$. For our purpose, we focus on the (010) surface theory which will be solved by replacing $k_y$ with $-i\partial_y$. 
The Hamiltonian can be then separated into two parts,
\begin{eqnarray}
&& H_{\perp} (-i\partial_y) = \nonumber \\ 
&& \begin{pmatrix}
-(M_2+C_2)\partial_y^2+(M_0+g_z) & 0 & 0 & -v\partial_y \\
0 & (M_2-C_2)\partial_y^2-(M_0-g_z) & -v\partial_y & 0 \\
0 & v\partial_y & -(M_2+C_2)\partial_y^2+(M_0-g_z) & 0 \\
v\partial_y & 0 & 0 & (M_2-C_2)\partial_y^2-(M_0+g_z) \\
\end{pmatrix}, \nonumber \\
&&
\end{eqnarray} 
and 
\begin{eqnarray}
H_\parallel (k_x, k_z)= \begin{pmatrix}
M_1k_z^2 + M_2k_x^2 & v_z k_z & 0 & vk_x \\
v_zk_z & -(M_1k_z^2 + M_2k_x^2) & vk_x & 0 \\
0 & vk_x & M_1k_z^2 + M_2k_x^2 & -v_zk_z \\
vk_x & 0 & -v_zk_z & -(M_1k_z^2 + M_2k_x^2) \\
\end{pmatrix},
\end{eqnarray}
In particular, $H_\perp$ consists of two $2$ by $2$ blocks,
\bea
h_{14}=\begin{pmatrix}
	-(M_2+C_2)\partial_y^2+(M_0+g_z)  & -v\partial_y \\
	v\partial_y & (M_2-C_2)\partial_y^2-(M_0+g_z) \\
\end{pmatrix},
\eea
and 
\bea
h_{23}=\begin{pmatrix}
	(M_2-C_2)\partial_y^2-(M_0-g_z)  & -v\partial_y \\
	v\partial_y & -(M_2+C_2)\partial_y^2+(M_0-g_z) \\
\end{pmatrix}.
\eea

\subsection{Surface State Solution}
We note that $h_{23}$ is exactly $h_{14}$ if we flip the sign of $M_0$ and $M_2$. Thus, we only need to solve for the surface state of $h_{14}$. Let us define $m_g = M_0+g_z$ and consider a trial surface state solution at an energy $E$,
\bea
\psi = \begin{pmatrix}
	A \\
	B \\
\end{pmatrix} e^{-\lambda y}
, \ \ 
h_{14} \psi = E \psi.
\eea
This leads to 
\bea
\begin{pmatrix}
	m_g-(M_2+C_2)\lambda^2-E  & v\lambda \\
	-v\lambda & -m_g+(M_2-C_2)\lambda^2-E \\
\end{pmatrix}
\begin{pmatrix}
	A \\
	B \\
\end{pmatrix} = 0.
\label{Eq: surface equation}
\eea
Therefore, we have
\bea
\psi_E = \begin{pmatrix}
	-m_g+(M_2-C_2)\lambda^2-E \\
	v\lambda \\
\end{pmatrix} e^{-\lambda y}
\eea
For a given energy $E$, we notice that Eq. \ref{Eq: surface equation} implies the existence of two decay soluations $\lambda_{1,2}$. Therefore, $\psi_E$ should be modified to
\bea
\psi_E = {\cal N}_1 \begin{pmatrix}
	-m_g+(M_2-C_2)\lambda_1^2-E \\
	v\lambda_1 \\
\end{pmatrix} e^{-\lambda_1  y} + 
{\cal N}_2 \begin{pmatrix}
	-m_g+(M_2-C_2)\lambda_2^2-E \\
	v\lambda_2 \\
\end{pmatrix} e^{-\lambda_2  y}
\eea
where ${\cal N}_{1,2}$ are the normalization factors. The boundary condition at $y=0$ enforces the vanishing of the surface state wavefunction and we find that
\bea
\begin{pmatrix}
	-m_g+(M_2-C_2)\lambda_1^2-E & -m_g+(M_2-C_2)\lambda_2^2-E \\
	v\lambda_1 & v\lambda_2 \\
\end{pmatrix}
\begin{pmatrix}
	{\cal N}_1 \\
	{\cal N}_2
\end{pmatrix} = 0
\eea
This immediately leads to
\bea
\lambda_1 \lambda_2 = -\frac{E+m_g}{M_2-C_2}. 
\label{Eq: l1 l2}
\eea
On the other hand, following Eq. \ref{Eq: surface equation}, we have
\begin{equation}
(C_2^2-M_2^2) \lambda^4 + (v^2+2M_2m_g + 2C_2E) \lambda^2 + (E^2-m_g^2) = 0
\label{Eq: l quartic equation}
\end{equation}
and consequently
\bea
\lambda_1 \lambda_2 = \frac{E^2-m_g^2}{C_2^2-M_2^2}. 
\label{Eq: l1 l2 number 2}
\eea
Together, since $\lambda_1\lambda_2\neq 0$, we have
\begin{equation}
E_{14} = -\frac{(M_0+g_z)C_1}{M_2}.
\end{equation}
Plug it into Eq. \ref{Eq: l quartic equation} and complete the square, we arrive at
\bea
(\lambda^2 - \frac{m_g}{M_2})^2 = - \frac{(v\lambda)^2}{C_1^2-M_2^2}.
\eea
and eventually we find that the spinor part of $\psi_E$ is
\bea
\xi_{14} = \begin{pmatrix}
	\sqrt{\frac{M_2-C_1}{M_2+C_1}} \\
	1 \\
\end{pmatrix}
\label{Eq: xi14}
\eea
Similarly, one can obtain 
\bea
E_{23} = \frac{(-M_0+g_z)C_1}{M_2},\ \ \xi_{23} = \begin{pmatrix}
	\sqrt{\frac{M_2+C_1}{M_2-C_1}} \\
	-1 \\
\end{pmatrix}
\label{Eq: xi23}
\eea

\subsection{Surface Hamiltonian and the Gap}
With the surface states in Eq. \ref{Eq: xi14} and Eq. \ref{Eq: xi23}, the projection of $H_{\parallel}$ is straightforward. Up to ${\cal O}(k)$ and ignoring the identity term, we have
\begin{equation}
H_{ss} = \tilde{v} k_x\sigma_z + \tilde{v}_z k_z \sigma_x - \frac{g_zC_1}{M_2}\sigma_z.
\end{equation}
Therefore, with out-of-plane FM, the surface Dirac point of (010) surface only gets shifted from the origin of surface BZ to $k_x=\frac{g_zC_1}{\tilde{v}M_2}$. The surface gap opens up only when we include the hexagonal warping effect. We find that $H_{ss}$ is modified to
\begin{equation}
H_{ss} = \tilde{v} k_x\sigma_z + \tilde{v}_z k_z \sigma_x - \frac{g_zC_1}{M_2}\sigma_z + \tilde{w} k_x^3\sigma_y.
\end{equation}
Therefore, {\bf the surface gap of (010) surface is a combined efect of out-of-plane FM $g_z$, the particle-hole breaking $C_1$, and the hexagonal warping $w$.} Since the momentum part of the hexagonal warping term is invariant under $C_{3z}$ but odd under inversion, it is expected that the hinge between neighboring surfaces on the hexagonal prism [in Fig. 4(c) of the main text] forms a mass domain wall for 2d Dirac fermions, which explains the existence of chiral hinge modes. 

\subsection{Surface Transition between HOTI $\alpha$ and $\beta$ Phases}
When the FM moment is rotated by the magnetic field and has an in-plane component, the surface topological transition between $\alpha$ and $\beta$ phases can be induced. In the bulk Hamiltonian, we consider a y-directional Zeeman term $g_y s_y\otimes \sigma_0$. Upon projection, we arrive at
\bea
H_{ss} = \tilde{v} k_x\sigma_z + \tilde{v}_z k_z \sigma_x - \frac{gC_1}{M_2}\sigma_z + (\tilde{w} k_x^3 - \tilde{g}_y )\sigma_y.
\eea
The transition happens when the surface gap closes. This is only possible when 
\bea
\tilde{g}_y = \tilde{w} k_x^3 = \frac{C_1^3 \tilde{w} }{\tilde{v}^3M_2^3} g_z^3.
\eea

\section{Appendix E: HOTI $\beta$ Phase with Canted AFM}

\begin{figure}[t]
	\centering
	\includegraphics[width=0.95\textwidth]{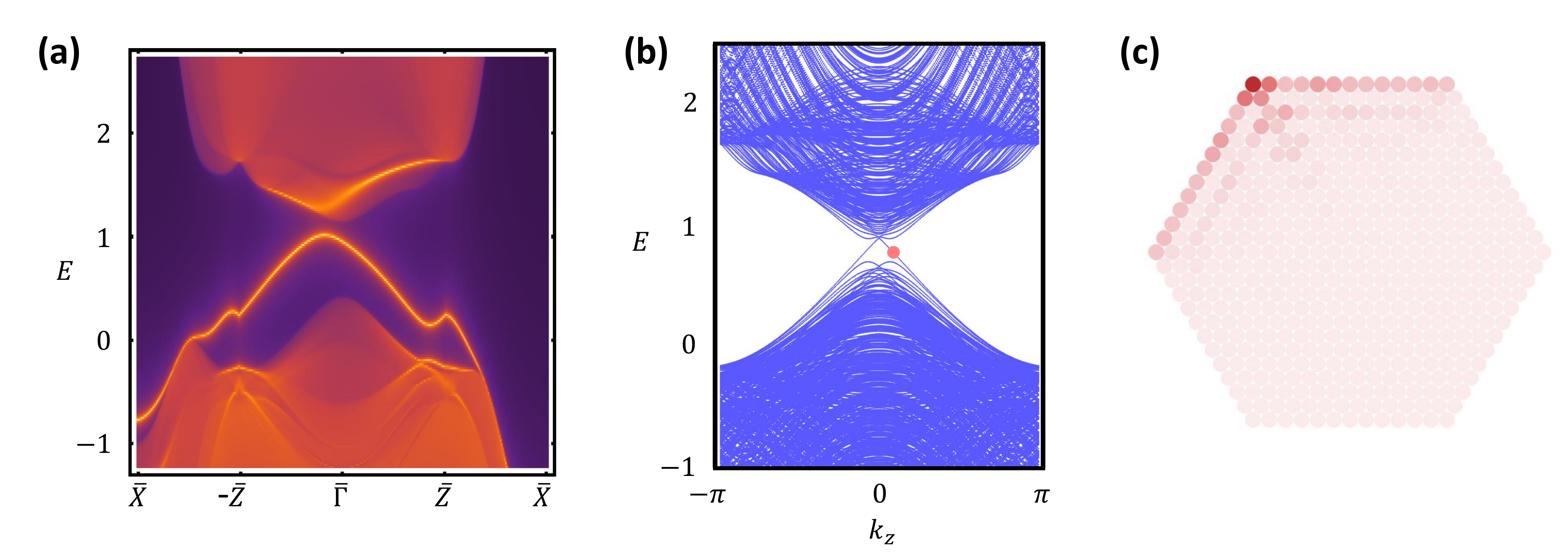}
	\caption{(a) The (010) surface spectrum calculated using iterative Green function method. (b) The energy spectrum in a prism geometry as a function of $k_z$. (c) The wavefunction distribution of the left-moving mode shown by the red dot in (b).}
	\label{Fig: canted AFM HOTI}
\end{figure}

In this section, we provide an example of HOTI $\beta$ phase with canted AFM ordering. The canted AFM we consider is characterized by $\phi_A = \phi_B = \pi /6$, $\theta_A = 0.3\pi$ and $\theta_B = 0.7\pi$, which explicitly breaks both $\Theta_M$, $C_{3z}$, $M_x$ and the glide mirror ${\cal G}_x$. The spatial inversion ${\cal I}$, however, remains preserved. 

The surface dispersion on the (010) surface is calculated and shown in Fig. \ref{Fig: canted AFM HOTI} (a). Just as we expect, the surface spectrum displays a finite energy gap due to symmetry breaking. We then calculate the energy spectrum in the prism geometry with the periodic boundary condition applied along $z$ direction. Inside the surface gap, there exists a pair of spatially-separated counterpropagating hinge modes that are locally chiral. For example, the spatial profile of the left-moving chiral channel is plotted in Fig. \ref{Fig: canted AFM HOTI} (c). 

The results in Fig. \ref{Fig: canted AFM HOTI} unambiguously establish this phase as a HOTI $\beta$ phase defined in the main text.  

\section{Appendix F: Macrospin Approximation for Field-Induced Canted AFM}
In the AFM phase of MnBi$_{2n}$Te$_{3n+1}$, the easy axis of the magnetic moment is along the out-of-plane $\hat{z}$ direction. When an external magnetic field ${\bf B}$ is applied perpendicular to this easy axis, the magnetic moments cant towards the field direction. This canting effect can be described by the Stoner-Wohlfarth model \cite{baltz2018antiferro}. In the macrospin approximation, the magnetic energy per magnetic unit cell is given by
\begin{equation}
\mathcal{E} = 2 J_{AF} {\bf M}_A \cdot {\bf M}_B - K ( M_{A, z}^2 + M_{B, z}^2) - g ({\bf M}_A + {\bf M}_B)\cdot {\bf B},
\end{equation} 
where ${\bf M}_{A, B}$ are unit vectors representing the direction of magnetic moments, $J_{AF}>0$ is the antiferromagnetic exchange coupling, $K>0$ is the uniaxial anisotropy, and $g$ is the effective Zeeman coupling. 

When ${\bf B}$ field is along in-plane $\hat{x}$ direction, ${\bf M}_A$ and ${\bf M}_B$ build up equal $\hat{x}$  component. The energy density $\mathcal{E}$ can be minimized by the canted AFM ansatz ${\bf M}_A = (n_x, 0, \sqrt{1-n_x^2})$ and ${\bf M}_B = (n_x, 0, -\sqrt{1-n_x^2})$, where $|n_x|<1$. In terms of $n_x$, $\mathcal{E}$ is given by:
\begin{equation}
\mathcal{E} = 2(2 J_{AF} + K) n_x^2 - 2 g n_x B_x - 2 (J_{AF} + K).
\end{equation} 
For a small in-plane ${\bf B}$ field, $\mathcal{E}$ is minimized by
\begin{equation}
n_x = \frac{g}{ 2(2 J_{AF} + K)} B_x.
\end{equation}
Therefore, the in-plane moments grow linearly with the in-plane field before saturating at $B_x^* =  2(2 J_{AF} + K)/g$. When $B_x$ exceeds $B_x^*$, the canted AFM phase is transformed to the in-plane ferromagnetic phase, as schematically plotted in Fig. 1(a) of the main text.

\end{document}